\documentclass{article}
\usepackage{arxiv}
\usepackage[utf8]{inputenc} 
\usepackage[T1]{fontenc}    
\usepackage{hyperref}       
\usepackage{url}         
\usepackage{booktabs}    
\usepackage{amsfonts}      
\usepackage{nicefrac}      
\usepackage{microtype}     
\usepackage{lipsum}
\usepackage{float}
\usepackage{graphicx}
\usepackage{caption}
\usepackage{subcaption}
\usepackage{titlesec}
\usepackage{multirow}
\usepackage{amsmath}
\usepackage{soul,color}
\usepackage{float}

\usepackage[section]{placeins}

\usepackage[nodisplayskipstretch]{setspace}

\title{DeepXRD, a Deep Learning Model for Predicting of XRD spectrum from Materials Composition}
\author{
  Rongzhi Dong \\
 Department of Computer Science and Engineering\\
  University of South Carolina\\
  Columbia, SC 29201 \\
    \And
 Yong Zhao\\
 Department of Computer Science and Engineering\\
  University of South Carolina\\
  Columbia, SC 29201 \\
  \And
 Yuqi Song, Nihang Fu, Sadman Sadeed Omee, Sourin Dey, Qinyang Li, Lai Wei\\
 Department of Computer Science and Engineering\\
  University of South Carolina\\
  Columbia, SC 29201 \\ 
  \And
 Jianjun Hu *\\
 Department of Computer Science and Engineering\\
  University of South Carolina\\
  Columbia, SC 29201 \\
  \texttt{jianjunh@cse.sc.edu} \\
}
\titlespacing {\section}
{0pt}{5.5ex plus 1ex minus .24ex}{3.3ex plus .24ex}
\begin{document}
\maketitle
\begin{abstract}
One of the long-standing problems in materials science is how to predict a material's structure and then its properties given only its composition. Experimental characterization of crystal structures has been widely used for structure determination, which is however too expensive for high-throughput screening. At the same time, directly predicting crystal structures from compositions remains a challenging unsolved problem. Herein we propose a deep learning algorithm for predicting the XRD spectrum given only the composition of a material, which can then be used to infer key structural features for downstream structural analysis such as crystal system or space group classification or crystal lattice parameter determination or materials property predictions. Benchmark studies on two datasets show that our DeepXRD algorithm can achieve good performance for XRD prediction as evaluated over our test sets. It can thus be used in high-throughput screening in the huge materials composition space for new materials discovery.

\end{abstract}

\section{Introduction}

\begin{figure}[ht]
  \centering
  \includegraphics[width=0.6\linewidth]{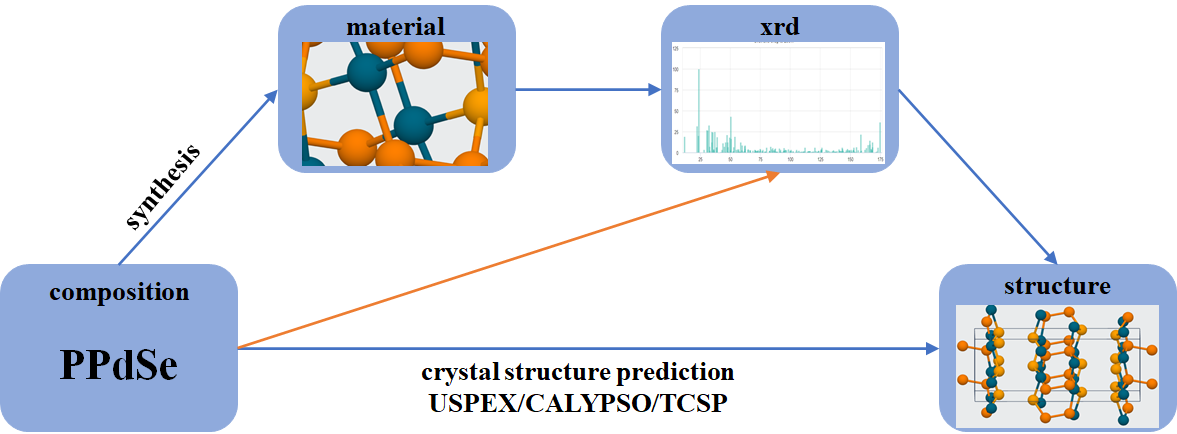}
  \caption{The composition-processing-structure-property relationships in materials science research. Experimental methods uses synthesis, XRD characterization to obtain the crystal structures while crystal structure prediction algorithms aim to directly predict the structures from compositions. This work tries the third approach: predicting the XRD spectrum from compositions.}
  \label{fig:paradigm}
\end{figure}

One of the major goals of materials science is to elucidate the composition-processing-structure-property-performance relationships so that materials with desired functions can be designed and synthesized \cite{hattrick2016perspective}. Traditionally, the problem is studied as a forward problem in which the cause-and-effect relationships are uncovered from composition to processing and structure and then to performance \cite{arroyave2019systems}. One starts with a tentative composition/recipe and then utilizes some known processing processes with adjustments to synthesize the material sample, whose structure is then derived using the structural characterization data, which is typically generated via scanning x-ray diffraction (XRD) or Raman spectroscopy experiments. By analyzing the structural characteristics, one can estimate its potential properties and performance. On the other hand, the materials discovery can be formulated as an inverse design problem, in which one starts from a performance target and try to find/search for the best composition and processing to achieve the desired performance. In both processes, one of the major bottlenecks is how to get the structure for a given composition. As shown in Figure \ref{fig:paradigm}, currently the experimental approaches are infeasible for large-scale screening of the vast chemical design space in which millions of possible compositions may be generated by modern generative models \cite{dan2020generative}. On the other hand, the computational crystal structure prediction algorithms such as USPEX, CALYPSO \cite{glass2006uspex, wang2012calypso} can only be applied to relatively small systems. The template-based crystal structure prediction methods such as \cite{wei2021tcsp} and \cite{kusaba2022crystal} are limited to predicting structures with known structure prototypes. 
In this paper, we aim to explore whether we can develop a deep learning algorithm to predict the XRD spectrum from the composition alone, which can then be used to characterize the structural properties of the materials such as geometric symmetry in terms of crystal systems and space groups \cite{suzuki2020symmetry}. XRD has also been used directly for unsupervised learning-based discovery of new superionic conductors \cite{zhang2019unsupervised} or as materials structural representations \cite{banko2021deep}. High quality predicted XRD spectrum may also be used to reconstruct crystal structures using well-established methods in the crystallography community \cite{harris2001contemporary}.

In order to study the crystal structures of inorganic materials, the first principles calculations such as Density Functional Theory (DFT) have been widely used for crystal structure prediction \cite{sikam2019study,khalid2019synthesis}. Although the first principles calculations are powerful, they are susceptible to the constraints of their excessive calculation cost, which limits the size of the material design space or the number of materials they can screen. To address this problem, machine learning (ML) has been increasingly applied to materials science fields, leading to the emergence of “materials informatics” \cite{rajan2005materials}, in which materials learning methods are developed to obtain prior knowledge and predictive models from known material datasets, and then predict complex material properties-based on these models. In the past few years, ML has succeeded in predicting new features \cite{ward2017atomistic}, guiding chemical synthesis, and discovering suitable compounds with target properties \cite{gomez2018automatic,popova2018deep,lu2018accelerated,collins2016materials}.

Several composition-based machine learning models have been proposed to predict structural properties such as crystal systems \cite{liang2020cryspnet}, space groups\cite{zhao2020machine}, or lattice constants\cite{takahashi2017descriptors}, with varying performances. Composition-based ML models have also been extensively used for material property prediction. Well-known composition descriptors such as Magpie\cite{ward2016general}, Matminer\cite{ward2018matminer}, composition-graph based embeddings\cite{goodall2019predicting} have all been proposed for structure or property prediction. While these composition-based ML models for such tasks have been criticized for lack of high performance compared to structural descriptors-based materials property prediction models, they have a unique advantage for de novo discovery of new materials of which the crystal structures are usually not available and then only composition-based ML models can be used \cite{dan2020generative}. In addition, such models can be used as the first level coarse screening of millions of generated hypothetical materials from generative machine learning models \cite{dan2020generative}. 

Several recent works have applied machine learning to XRD data. Suzuki et al. \cite{suzuki2018machine} used Random forest models to predict crystal systems and 230 space groups from XRD. A deep learning approach has also been proposed for space group classification from XRD\cite{park2017classification}. Oviedo et al. \cite{oviedo2019fast} proposed a physics-informed data augmentation method that extends small, targeted experimental and simulated datasets and developed a convolutional neural network for classifying seven space groups. Convolutional neural networks have also been used to map the XRD patterns to materials with one-to-one mapping\cite{wang2020rapid}. More recently, a deep neural network model \cite{kaufmann2020crystal} has been reported in \textit{Science} to autonomously identify the crystal symmetry (systems) from Electron backscatter diffraction and achieved high accuracy. 
Another related work is the XRD-based phase attribution or phase diagram reconstruction\cite{xiong2017automated, li2018inferring}. In a typical XRD spectrum, the most critical information of the structure is encoded in the peak positions and corresponding peak height. In many cases, small peak shifts may also happen. In addition, the XRD spectrum has been shown to be used to predict the space group of the sample with high accuracy. XRD spectrum has also been used as a feature for unsupervised clustering to find new lithium superionic materials\cite{zhang2019unsupervised}.

Instead of trying to reconstruct the 3D coordinates of the crystal structure using DFT-based evolutionary search algorithms which are feasible only for small systems, in this work, we aim at emulating the traditional XRD-based structure characterizing approach: building a deep learning-based prediction model to predict the XRD given its composition. Experimentalists have been using XRD to analyze materials' properties for a long time. XRD predicted by our models can then be used by them for quick downstream analysis such as structure determination or property prediction. For example, XRD diffraction patterns have been used to achieve over 90\% accuracy for crystal system classification, except for triclinic cases, and with 88\% accuracy for space group classification with five candidates\cite{suzuki2020symmetry}. In \cite{vecsei2019neural}, neural networks were used for space group classification with an accuracy of around 54\% on experimental data, which however can be improved to 82\% at the cost of having half of the experimental data unclassified.

In this paper, we aim to develop deep learning based models to simulate the relationship between materials composition and XRD spectrum with the understanding that many chemically and structurally similar materials share close XRD spectra. 

Our contributions can be summarized as follows:
\begin{itemize}
    \item We develop two benchmark datasets for the composition based XRD prediction problems: $ABC_3$-XRD with 4270 samples and the Ternary-XRD dataset with 37,211 samples.
    \item We propose a deep learning-based neural network model for predicting XRD spectra from material compositions. 
    \item We evaluate four different loss functions based on different distance measures for calculating XRD similarities and find that the Pearson loss function achieves the best result. We find Mean Square Error (MSE) is not a good choice for training deep learning models for XRD prediction due to their sensitivity to the peak intensities. 
    \item We conduct extensive experiments over the two datasets and show that our proposed framework is capable to achieve good performance for test sets.
\end{itemize}

The remainder of this paper is organized as follows. Section 2 focuses on the research framework, materials representation, and evaluation indicators. Section 3 describes our experiments and highlights our prediction performance. The last section concludes the paper.

\section{Materials and Methods}

\subsection{XRD spectrum prediction problem}

In our composition-to-XRD mapping problem, the goal is to design a model that could learn from inorganic materials composition and then predict their probable XRD spectrum. We prepare two datasets for training our models. The smaller dataset has 4270 different inorganic materials with the prototype of $ABC_3$, where A, B, C are three different elements. The larger dataset has 43,273 samples of ternary materials. Each independent material has a corresponding XRD spectrum. In the case of polymorphism where one composition corresponds to multiple phases, we pick the structure with the lowest formation energy. According to the composition of a material, we need to predict what the XRD spectrum is. To evaluate our model performance, for each dataset, we randomly select 20\% as the test set from all samples 70\% as the training set, and 10\% as the validation set. Training set is used to train our prediction model and use validation set to tune the hyper-parameters. Finally, for a given target formula, we use our model to predict its XRD and compare it with the true XRD.

\begin{figure}[ht]
  \centering
  \includegraphics[width=1.0\linewidth]{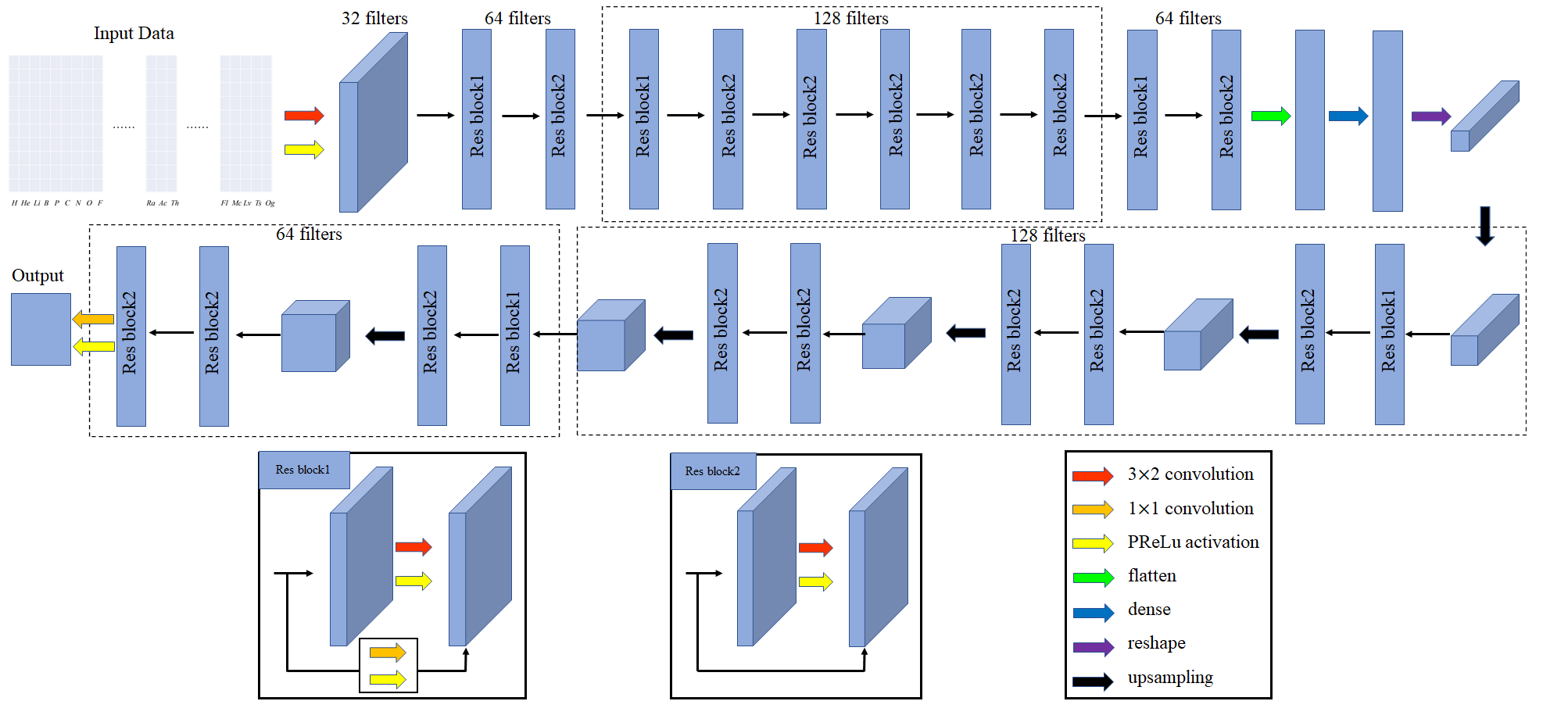}
  \caption{The XRD spectrum prediction framework. The input is the one-hot encoding of a given formula, through the first row's encoder to extract the key information and then follow the steps shown in the second row to decode and reconstruct the given material's probable XRD spectrum which is our output. To apply skip connections on layers with filters of different sizes and the same sizes, we use two different kinds of residual blocks with their structures shown in the left and middle bottom. To connect layers with different filters, we add convolution operations when using skip connections to make sure the two layers keep the same shape so that they can be added together, as shown in res block1. For layers with the same filters, we can just skip connections and add them together as shown in res block2. We use colorful arrows to represent different operations in our prediction model as shown in the bottom right.}
  \label{fig:framework}
\end{figure}

The main components of our deep learning framework are shown in Figure \ref{fig:framework}. We use deep residual network (ResNet) \cite{he2016deep} model trained with one-hot composition encoding features to learn the relationship between material composition and XRD spectrum. For a given material formula, we use its one-hot matrix as DeepXRD model's input. Due to one-hot matrix is sparse, as shown in the first row of Figure \ref{fig:framework}, after the first convolution layer, we use several ResNet blocks with different filters, flatten layer, and dense layer as encoding part to abstract the key information of input matrix. To reconstruct XRD spectrum, we use several upsampling layers to magnify the key information to a 32$\times$32 matrix as our output, this progress is shown in the second row of Figure \ref{fig:framework}. The last row shows two different ResNet block structures. For skip connection, the two inputs must have the same shape, if the two layers have different filter numbers, we must add a convolution step to make the two inputs keep with the same shape, as Res block1 shown in left bottom of Figure \ref{fig:framework}, and if the two layers have same filter number, we can just use skip connection as Res block2 at the middle bottom. We introduced all operations used in our model in the right bottom, we have a 3$\times$2 convolution layer, a 1$\times$1 convolution layer, flatten layer, dense layer, reshape layer and upsampling layer, the activation function we used in our model is PReLu. Operations used in each layer are indicated by arrows of different colors.

\subsection{Materials representation}

\begin{figure}[ht]
  \centering
  \includegraphics[width=0.8\linewidth]{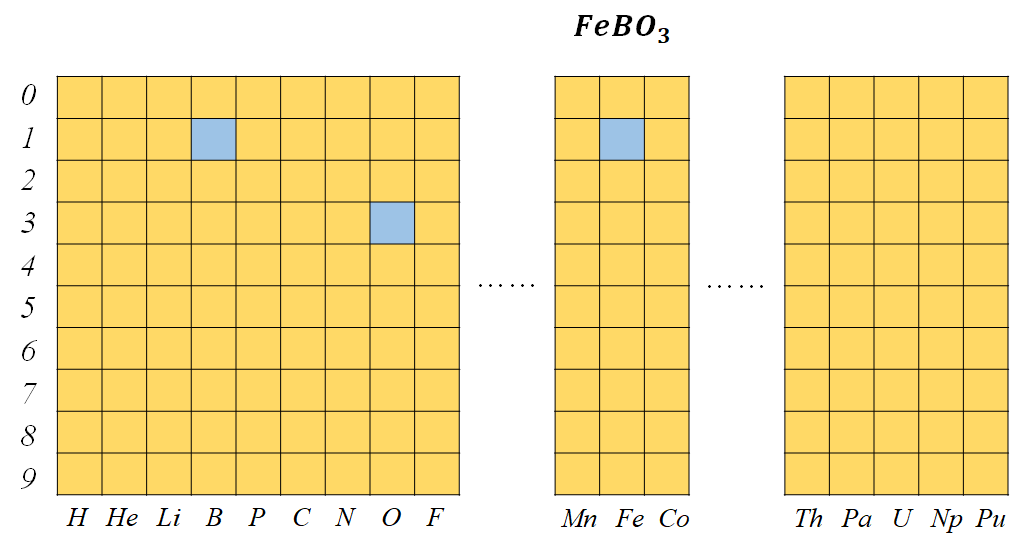}
  \caption{One-hot representation of formula $FeBO_3$, blue cells indicate 1 and yellow cells indicate 0.}
  \label{fig:onehot}
\end{figure}

We use the one-hot representation of the formula for XRD prediction, which has been used in \cite{dan2020generative}. The advantage of the one-hot encoding is that it can encode a discrete material's elemental composition with a discrete 2D matrix of binary values {0, 1}, which are extremely suitable for the convolution neural network layers to extract hierarchical patterns from it.
This coding method is also suitable for the characterization of the elements in the molecular formula of the material. As shown in Figure \ref{fig:onehot}, each formula can be encoded as a 2D matrix of dimension $N \times M$, where $N$ is the maximum number of atoms for an element in the formula and $M$ is the number of elements considered in our datasets. All elements not included in the material formula are set to 0, and the column corresponding to each element in the formula has a non-zero value of 1, which is assigned to the column cell in the row $j+1$  where $j$ is the number of atoms with this element in the material formula. For example, for $AcBO_3$, the one-hot code corresponding to this formula is a two-dimensional matrix with 10 rows and 84 columns (the samples in our datasets are composed of 64 different elements), in which the column Fe, B, and O  has value 1 on row 4, 2, and 2, respectively, and the remaining values of the matrix are all set to 0. 

\subsection{ResNet model for XRD prediction from composition}

\begin{figure}[ht]
  \centering
  \includegraphics[width=0.6\linewidth]{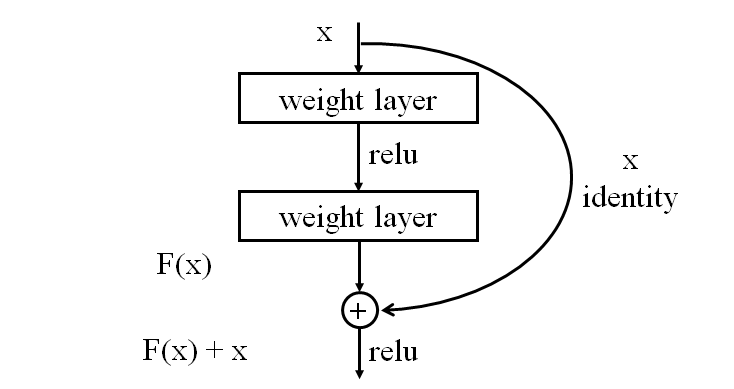}
  \caption{Building block of ResNet, introduced by \cite{he2016deep}.}
  \label{fig:resnet}
\end{figure}

Figure \ref{fig:framework} shows the whole neural network architecture of our DeepXRD model, which contains an input part, an encoding module, and a decoding module. The input data of our model is the one-hot representation of a given material formula with the dimension of 10$\times$84, and the output is a matrix that represents the corresponding XRD spectrum with a dimension 32$\times$32. Since the one-hot encoding matrix is very sparse, we use the encoding module to extract the key information of materials and then use the decoding part to reconstruct the XRD spectrum. Our network is mainly composed of two types of residual blocks. Our res blocks are shown in the bottom of Figure \ref{fig:framework}. Figure \ref{fig:resnet} shows the basic residual block with the shortcut connections. In convolutional neural networks, the output from the layer and the identity input may have different dimensions, so we add convolution operations in the shortcut connection such that the input is converted to the same dimensions. The res block1 is used to connect layers with different number of filters so we add convolution operations when making skip connections to make sure the two parts we add together have the same dimension. Res block2 is used for layers with the same number of filters, so we can just add them together. The use of residual blocks aims to address the vanishing gradient problem in training deep convolutional neural networks\cite{he2016deep}.

In our DeepXRD model we choose the Parametric Rectified Linear Unit (PReLu)\cite{he2015delving} instead of Rectified Linear Unit(ReLu) as the activation function. The basic ReLU has an output of 0 if the input is less than 0, which could cause the dying ReLU problem\cite{lu2019dying} where some ReLU neurons essentially remain inactive for all inputs. Due to the slope of ReLU in the negative part is also 0, once a neuron gets negative, it's unlikely to recover anymore. Therefore no gradient flows and a large part of neural networks may do nothing. Parametric ReLU (PReLU) has a small slope for negative input values, which fixes the dying ReLU problem and can also speed up training. PReLU improves model fitting with nearly zero extra computational cost and little overfitting risk\cite{he2015delving}. 
As shown in Figure \ref{fig:framework}, after the first convolution layer, there are 10 residual blocks with different numbers of filters, which reduce the feature map matrix size until 11$\times$2. Flatten, dense, and reshape layers are then used to convert the feature map matrix to 1$\times$1$\times$128. The decoding module uses upsampling layers and res blocks layers to increase the feature map matrix size to get the final XRD value matrix. We compare four different loss functions and finally choose the Pearson product-moment correlation as the loss function for our DeepXRD model. The parameters of each layer are shown in Table \ref{table:parameters}.

\begin{table}[H] 
\begin{center}
\caption{Layers and parameters of the DeepXRD model.}
\label{table:parameters}
\begin{tabular}{|l|l|l|l|l|l|}
\hline
\textbf{Layer} & \textbf{Input Shape} & \textbf{Filter} & \textbf{Layer} & \textbf{Input Shape} & \textbf{Filter}  \\ \hline
conv1            & {[}batch, 84,10,1{]}   & 32 & Res block1            & {[}batch, 2,2,128{]}     & 128\\ \hline
Res block1    & {[}batch, 84,10,32{]}& 64 &Res block2& {[}batch, 2,2,128{]} & 128  \\ \hline
Res block2            & {[}batch, 42,5,64{]}     & 64  &upsampling            & {[}batch, 2,2,128{]}     & 128\\ \hline
Res block1            & {[}batch, 42,5,64{]}     & 128 & Res block2            & {[}batch, 4,4,128{]}     & 128 \\ \hline
Res block2            & {[}batch, 21,3,128{]}    & 128 &Res block2            & {[}batch, 4,4,128{]}     & 128  \\ \hline
Res block2            & {[}batch, 21,3,128{]}    & 128 &upsampling            & {[}batch, 4,4,128{]}     & 128  \\ \hline
Res block2            & {[}batch, 21,3,128{]}    & 128 &Res block2            & {[}batch, 8,8,128{]}     & 128 \\ \hline
Res block2            & {[}batch, 21,3,128{]}    & 128 & Res block2            & {[}batch, 8,8,128{]}     & 128 \\ \hline
Res block2            & {[}batch, 21,3,128{]}    & 128  & upsampling            & {[}batch, 8,8,128{]}     & 128\\ \hline
Res block2            & {[}batch, 21,3,128{]}    & 64  & Res block1            & {[}batch, 16,16,64{]}     & 64\\ \hline
Res block2            & {[}batch, 11,2,64{]}     & 64 &Res block2            & {[}batch, 16,16,64{]}     & 64 \\ \hline
flatten               & {[}batch, 11,2,64{]}  &  &upsampling            & {[}batch, 16,16,64{]}     & 64\\ \hline
dense               & {[}batch, 1408{]}  &   &Res block1            & {[}batch, 32,32,64{]}     & 64\\ \hline
reshape               & {[}batch, 128{]}  &  &Res block2            & {[}batch, 32,32,64{]}     & 64 \\ \hline
upsampling            & {[}batch, 1,1,128{]}     & 128&conv                  & {[}batch, 32,32,64{]}     & 1  \\ \hline
\end{tabular}
\end{center}
\end{table}

\subsection{Materials datasets}

To evaluate the performance of our DeepXRD algorithm, we prepared two datasets. The first dataset, $ABC_3$-XRD, contains 4270 material compositions of the prototype $ABC_3$ along with the XRD spectra calculated for their crystal structures with the lowest formation energy as downloaded from the Materials Project database \cite{Jain2013}. Since all the materials share the same prototype except the elements, we expect that the algorithm will achieve better performance over this dataset. The second dataset Ternary-XRD contains 37,211 compositions of diverse prototypes along with their computed XRD as downloaded from the Materials Project database \cite{Jain2013}. Each of the XRD data contains the corresponding XRD intensity at a set of 2$\theta$(180) degrees. We used a Gaussian smoothing procedure to sample the raw XRD spectrum so that all XRD sample contains 1024 points ranging from 0 to 180 degrees. 

\begin{figure}[h!]
  \centering
  \includegraphics[width=0.4\linewidth]{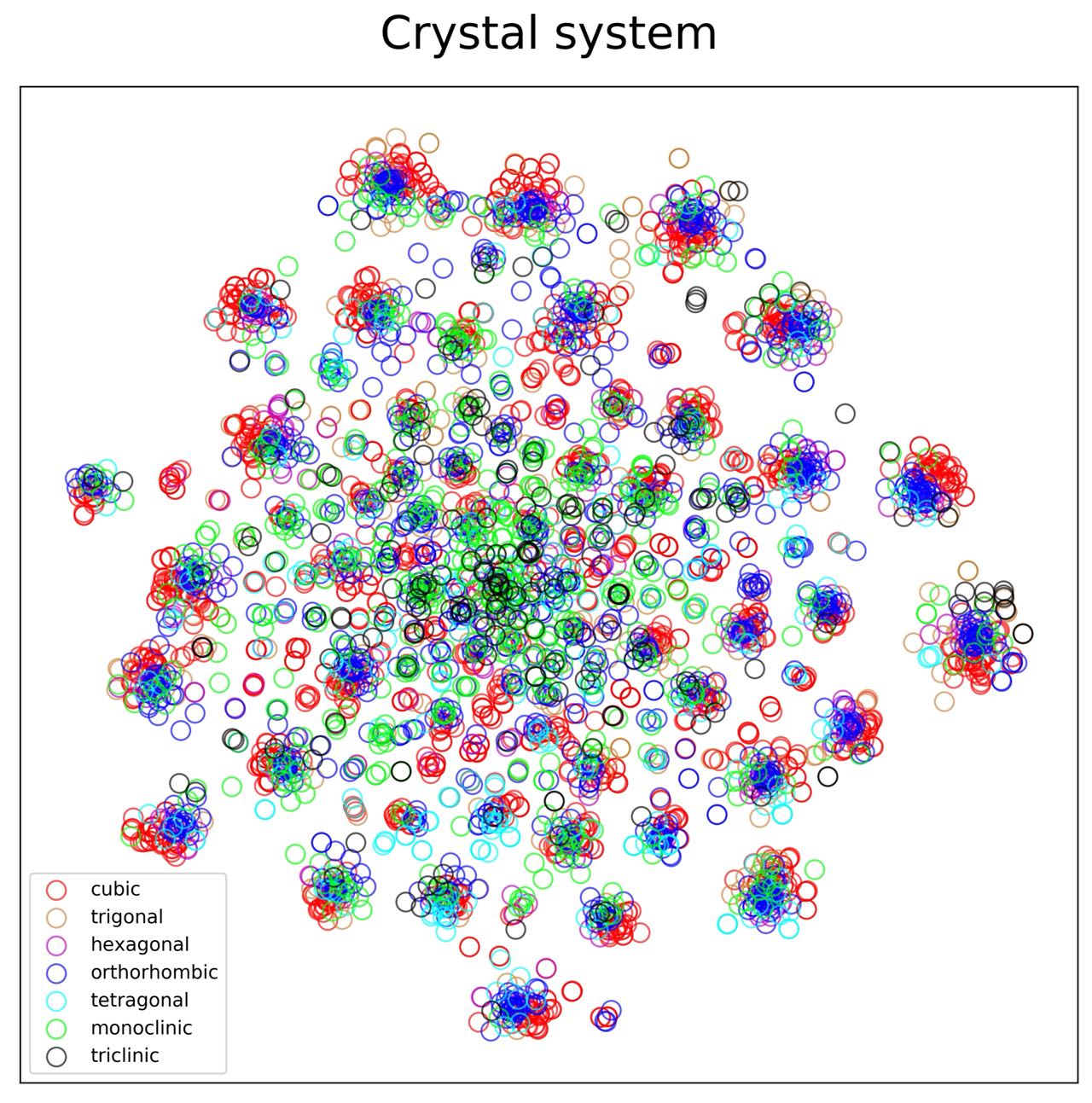}
  \caption{The XRD distribution of the $ABC_3$-XRD dataset using t-SNE visualization of XRD patterns. Each point corresponds to one XRD pattern. XRD patterns form loose clusters for each crystal system.}
  \label{fig:t-sne}
\end{figure}

In the crystallography community, it is well known that compositions in a given materials families tend to have similar XRD spectra. It's interesting to check whether XRD data also form clusters. We thus visualize the distribution of the $ABC_3$-XRD dataset using the t-distributed Stochastic Neighbor Embedding(t-SNE) \cite{JMLR:v9:vandermaaten08a} technique to map high-dimensional XRD spectra into a two-dimensional map, each point in Figure \ref{fig:t-sne} corresponding to one XRD spectrum. From the data distribution figure, we find that there are several loose cluster sets with the same crystal system which means these materials may also have similar structural of chemical properties.

\subsection{Model evaluation}

 \paragraph{Postprocesing: } there is always some noise in XRD caused by the impurities in the material.To ignore the influence of these noises, we set a threshold of XRD magnitude values to distinguish between the main components and the impurities. As the largest XRD magnitude value is 100 count per second (cps), we set 10cps as the threshold value: if the XRD magnitude values are smaller than 10cps, they are more likely caused by impurities rather than main components. Therefore we can just ignore them and consider values greater than 10cps as true peaks.
 We also introduce a peak-alignment operation to consider the allowable peak shifts in experimental XRD based structure characterization. In performance evaluation stage, we have corresponding ground truth XRD peaks so we can shift predicted XRD peaks within a threshold distance to do peak alignment: for a given ground truth peak, we first find all peaks within 2 degrees of peak position and then consider the largest peak as the corresponding peak of the true peak for prediction error calculation.  In testing stage, we do not have ground truth XRD value, so we can just do peak merging: as all peak positions range from 0 to 2$\theta$ (in our dataset, 2$\theta$ is 180 degrees), we can safely consider that there is just one peak if the distance between several peaks is less than 2 degrees. In this case, we assume that the main peak's position is at the middle of these peaks. By disregarding of noises and applying peak alignment, the predicted XRD values can be more realistic.  
 
To evaluate the performance of our DeepXRD, we introduce a series of XRD dissimilarity/distance metrics including Cosine matrix, Pearson product-moment correlation, Jensen–Shannon divergence (JSD), and Dynamic time warping (DTW), which have been evaluated in \cite{iwasaki2017comparison}. It is found that the Cosine and Pearson similarity measures can get the best XRD clustering performance when peak height changes and peak shifting are present in the data (due to lattice constant changes) and the magnitude of peak shifting is unknown. In another study \cite{hernandez2017}, 49 metrics are evaluated to check how sensitive they are when used for XRd clustering. It is shown when the prior knowledge of the maximum peak shifting is available, dynamic time warping in a normalized constrained mode provides the best clustering performance. For two diffraction patterns $\boldsymbol{s}$ and $\boldsymbol{t}$, the dissimilarity measure is defined as D($\boldsymbol{s}$, $\boldsymbol{t}$). For D($\boldsymbol{s}$, $\boldsymbol{t}$) = 0, the two diffraction patterns are assumed to be identical and the corresponding samples are assumed to share the same structure. Larger values of the dissimilarity measure imply greater dissimilarity between the samples’ structures.

The evaluation measures used in our work are shown as follows: we use MSE, 
mean squared logarithmic error(MSLE), Cosine metric, Pearson product-moment correlation, JSD, and DTW\cite{salvador2007toward} to compute XRD dissimilarity.

\begin{equation}
D_{\mathrm{MSE}}(\boldsymbol{s}, \boldsymbol{t})=\frac{1}{n}{\textstyle\sum_{i=1}^{n}}\left(s_{i}-t_{i}\right)^{2}
\end{equation}

\begin{equation}
D_{\mathrm{MSLE}}(\boldsymbol{s}, \boldsymbol{t})=\frac{1}{n}{\textstyle\sum_{i=1}^{n}}\left(\log\left(s_{i}+1\right)-\log\left(t_{i}+1\right)\right)^{2}
\end{equation}

\begin{equation}
D_{\text {Cosine }}(\boldsymbol{s}, \boldsymbol{t})=1-\frac{\sum_{i=1}^{n}\left(s_{i} \cdot t_{i}\right)}{\left(\sum_{i=1}^{n} s_{i}^{2}\right)^{\frac{1}{2}}\left(\sum_{i=1}^{n} t_{i}^{2}\right)^{\frac{1}{2}}}
\end{equation}

\begin{equation}
D_{\text {Pearson }}(\boldsymbol{s}, \boldsymbol{t})=1-\frac{\sum_{i=1}^{n}\left(s_{i}-\bar{s}\right)\left(t_{i}-\bar{t}\right)}{\left(\sum_{i=1}^{n}\left(s_{i}-\bar{s}\right)^{2}\right)^{\frac{1}{2}}\left(\sum_{i=1}^{n}\left(t_{i}-\bar{t}\right)^{2}\right)^{\frac{1}{2}}}
\end{equation}

\begin{equation}
D_{\mathrm{JSD}}(\boldsymbol{s}, \boldsymbol{t})=\frac{1}{2}{\textstyle\sum_{n}^{i=1}}s_{i}\cdot\log\left(\frac{2s_{i}}{s_{i}+t_{i}}\right)+\frac{1}{2}{\textstyle\sum_{n}^{i=1}}t_{i}\cdot\log\left(\frac{2t_{i}}{s_{i}+t_{i}}\right)
\end{equation}

\section{Results and discussion}

\subsection{Prediction performance of the composition descriptor-based XRD predictor}

\begin{figure}[ht!]
  \centering
  \begin{subfigure}{.24\textwidth}
    \includegraphics[width=\textwidth]{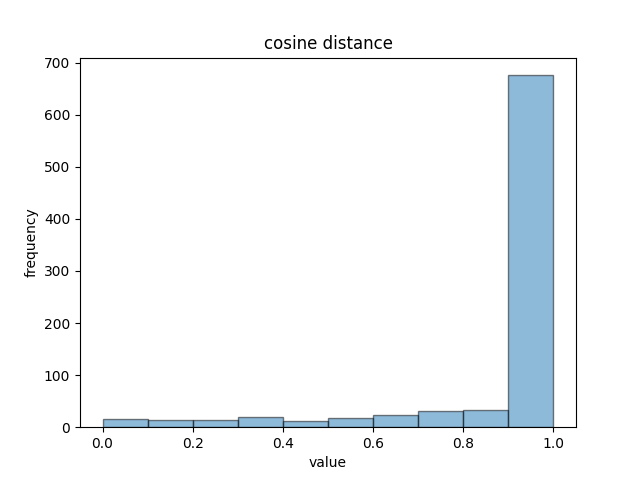}
    \vspace{3pt}
  \end{subfigure}
  \begin{subfigure}{.24\textwidth}
    \includegraphics[width=\textwidth]{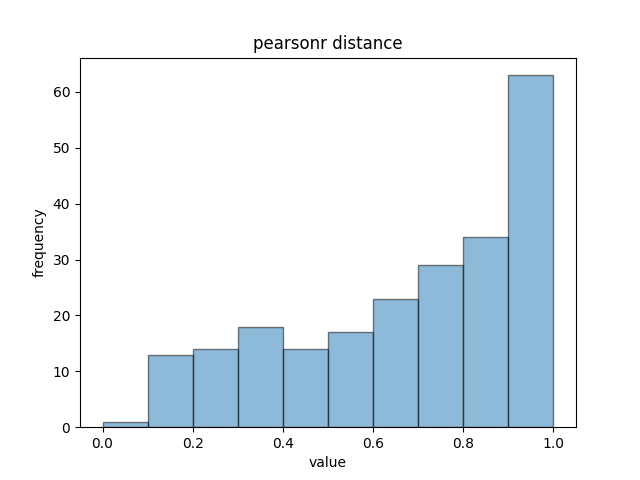}
    \vspace{3pt}
  \end{subfigure}
    \begin{subfigure}{.24\textwidth}
    \includegraphics[width=\textwidth]{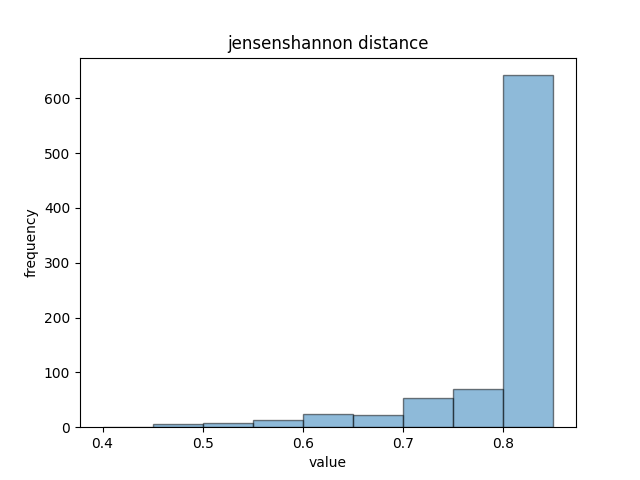}
    \vspace{3pt}
  \end{subfigure}
  \begin{subfigure}{.24\textwidth}
    \includegraphics[width=\textwidth]{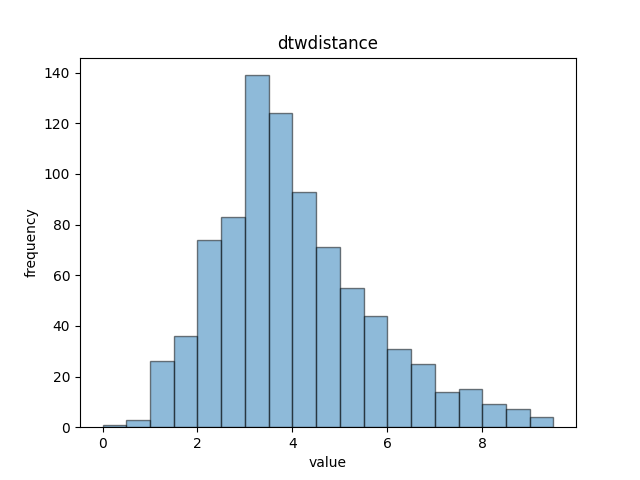}
    \vspace{3pt}
  \end{subfigure}
    \begin{subfigure}{.24\textwidth}
    \includegraphics[width=\textwidth]{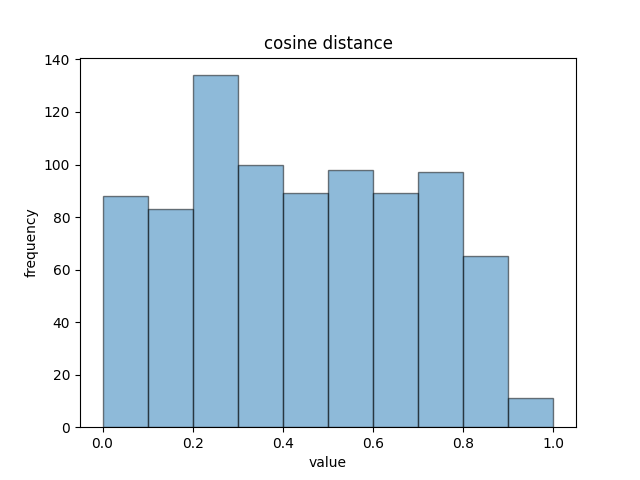}
    \vspace{3pt}
  \end{subfigure}
  \begin{subfigure}{.24\textwidth}
    \includegraphics[width=\textwidth]{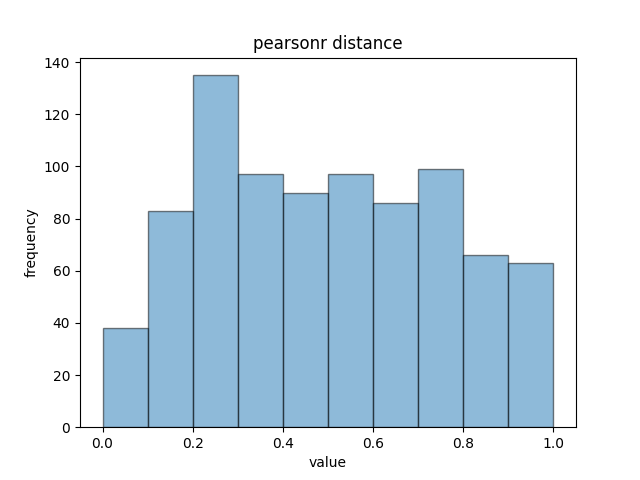}
    \vspace{3pt}
  \end{subfigure}
    \begin{subfigure}{.24\textwidth}
    \includegraphics[width=\textwidth]{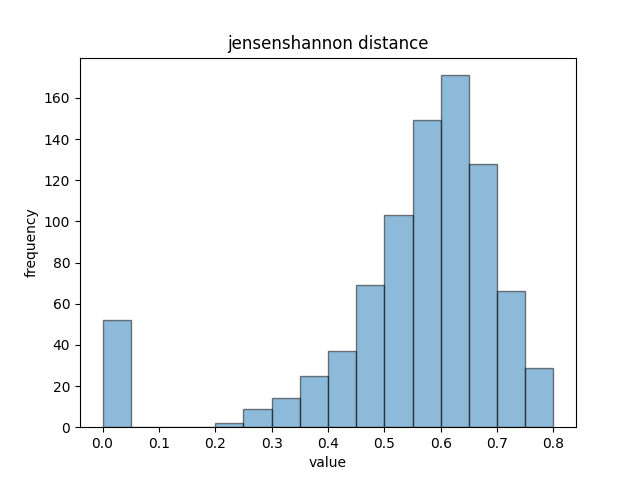}
    \vspace{3pt}
  \end{subfigure}
  \begin{subfigure}{.24\textwidth}
    \includegraphics[width=\textwidth]{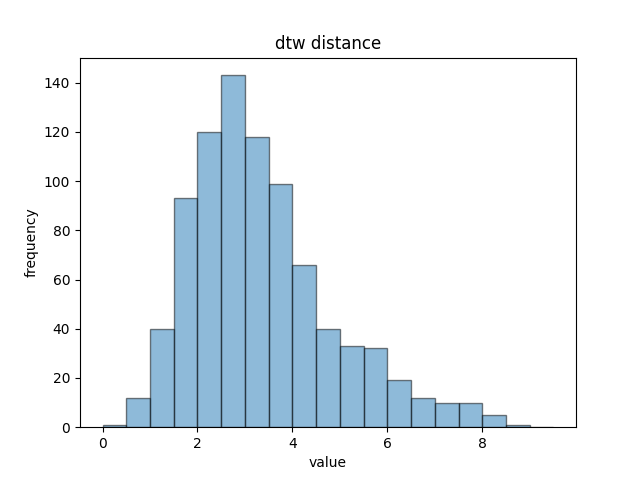}
    \vspace{3pt}
  \end{subfigure}
  \caption{Histogram of peak distance errors over the testing samples.}
    \label{fig:distance}
\end{figure}

In XRD based crystal structure characterization, it is the peak positions rather than their magnitude values that mainly reflect their structural or chemical property of the materials. Small XRD magnitude values are usually caused by impurities, and therefore independent of the material properties. Thus we can mainly focus on the peak positions of our predicted results. For each formula, after generating XRD values, we first select its peak position (where magnitude values are larger than 10cps) and then use peak merging to combine peaks if their position's distance is less than 2 degrees. Table \ref{table:distance} uses Cosine, Pearson, JSD, and DTW algorithms to evaluate mean distance errors of testing samples to show the predicted performance of the DeepXRD model. We use the Pearson as module's loss function. The first row is the distance between all predicted XRD position and target position, in the second row we only focus on the peak positions between predicted and target XRD spectra, in the last row we use peak merging to apply shifting on predicted peak position to make it clear. By focusing on peak position, the errors calculated by Cosine, Pearson, and JSD functions have increased from 0.884, 0.885, and 0.773 to 0.943, 0.950, and 0.811 respectively. The increase may caused by the accurate non-peak positions we ignored in this step. However, after merging peaks, the errors of these 3 functions have reduced to 0.633, 0.633, and 0.550 respectively. The decrease shows that although predicted peak positions are not totally exactly, most of them are very close to the ground truth positions. The errors calculated by DTW function are keep reducing from 4.54 to 3.92 and then 3.57. DTW can warping the sequence so it can calculate the distance of peaks of similar wave shape. Table \ref{table:distance} shows that DeepXRD model can find the key peak position of materials only through its composition. The distance errors distribution of all testing samples is shown in Figure \ref{fig:distance}. The first row shows the predicted peak position distance distribution and the second row are the distance errors after ignore noise and apply peak merging. The distance distribution figures show that by using peak merging, predicted XRD peak positions can be more similar to true peak positions thus the distribution of errors is more close to 0.

\begin{table}[ht!]
\begin{center}
\caption{Testing errors evaluated by 4 different performance measures for composition based XRD prediction.}
\label{table:distance}
\begin{tabular}{l|llll} 
\hline
  & Cosin        & Pearson      & JSD      & DTW         \\ \hline
prediction distance     & 0.884  & 0.885 & 0.773 & 4.54 \\ \hline
peak distance           & 0.943 & 0.950 & 0.811  & 3.92 \\ \hline
peak alignment distance & 0.633 & 0.633  & 0.550 & 3.57 \\ \hline
\end{tabular}
\end{center}
\end{table}

To choose the best loss function for the DeepXRD model, we use the peak match percentage as the criterion to evaluate four different loss functions' performance on the $ABC_3$-XRD and Ternary-XRD datasets respectively while keeping other hyper-parameters such as the batch size, learning rate and training epoch unchanged. We compare two traditional loss functions, MSE and MSLE with Cosine and Pearson, which have been shown to perform better in XRD similarity studies \cite{hernandez2017,iwasaki2017comparison}. Table \ref{table:loss} shows that all loss functions achieve better peak match percentages on the smaller $ABC_3$-XRD dataset. The peak match accuracy improves about 1\% compared with the larger Ternary-XRD dataset. We also find that the models trained with the Pearson loss function achieve the best match percentages on both datasets, which are 0.681 and 0.678 respectively. Compared with MSE's 0.626 and 0.612, MSLE's 0.644 and 0.631, Cosine's 0.673 and 0.667, the match percentages have improved 6\%, 4\%, and 1\% respectively. The results of Table \ref{table:loss} proves that for the XRD prediction problem, Cosine and Pearson loss functions are better than traditional MSE and MSLE loss functions because they focus more on the shape rather than the exact values.

\begin{table}[ht!]
\begin{center}
\caption{Prediction performance (peak position match percentage) of different loss functions.}
\label{table:loss}
\begin{tabular}{l|lllll}
\hline
 &         \multicolumn{4}{c}{performance measure}   \\ \hline
Dataset & MSE   & MSLE  & Cosine & Pearson  \\ \hline
$ABC_3$-XRD    & 0.626 & 0.644 & 0.673  & \textbf{0.681}    \\ \hline
Ternary-XRD & 0.612 & 0.631 & 0.667  & \textbf{0.678}    \\ \hline
\end{tabular}
\end{center}
\end{table}

\begin{figure}[h]
  \centering
  \begin{subfigure}{.33\textwidth}
    \includegraphics[width=\textwidth]{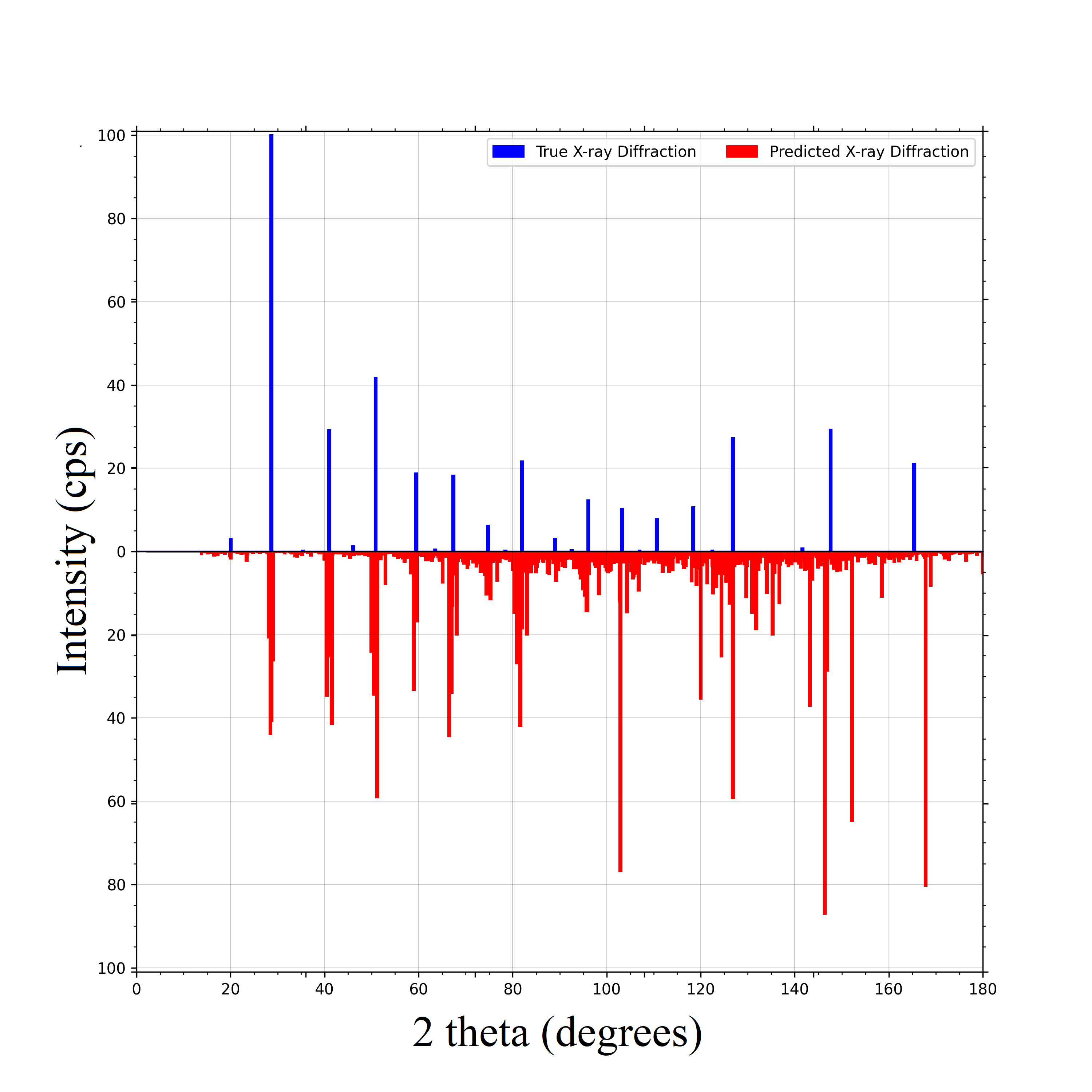}
    \caption{predicted XRD VS ground truth}
    \vspace{3pt}
  \end{subfigure}
  \begin{subfigure}{.33\textwidth}
    \includegraphics[width=\textwidth]{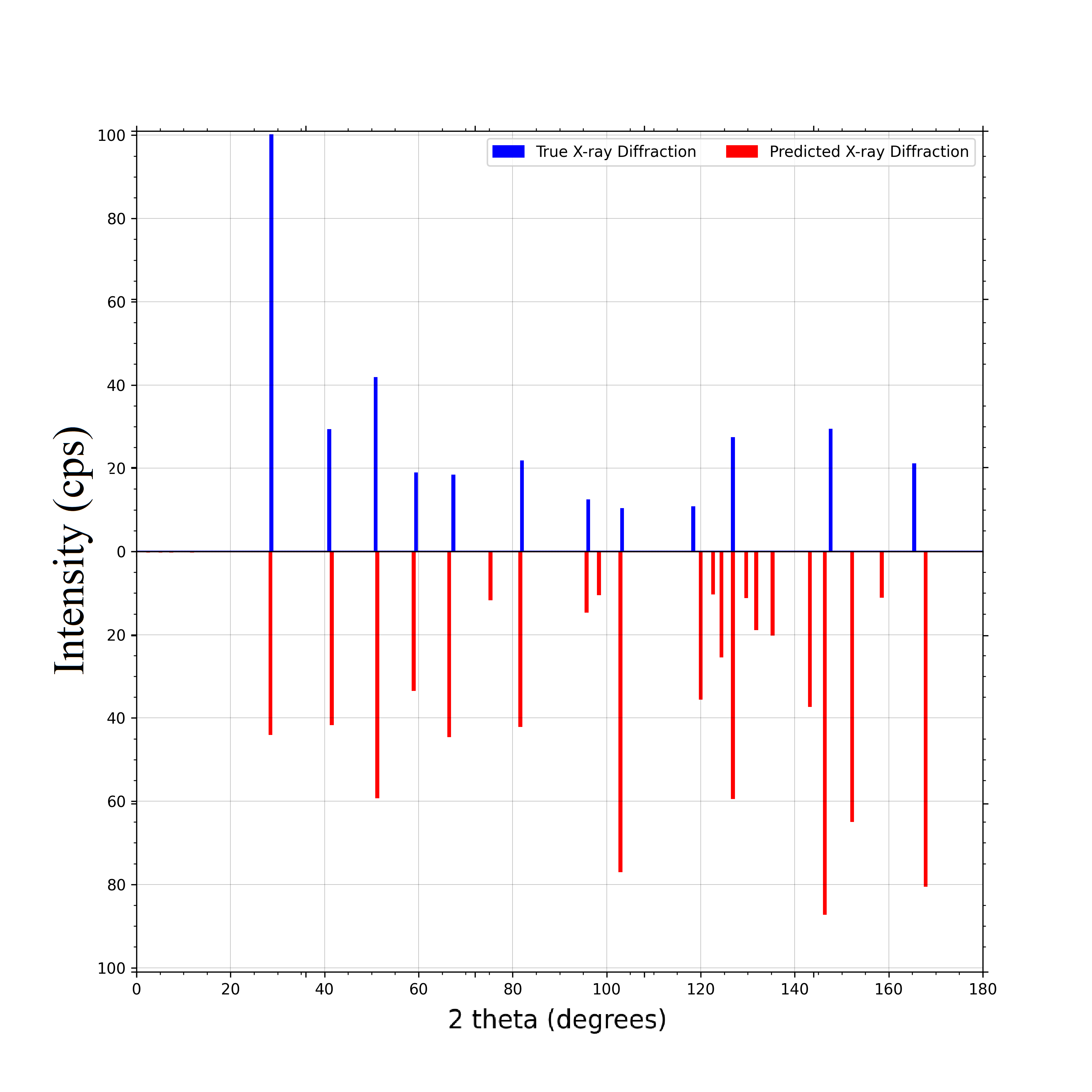}
    \caption{XRD peak VS ground truth}
    \vspace{3pt}
  \end{subfigure}
  \begin{subfigure}{.33\textwidth}
    \includegraphics[width=\textwidth]{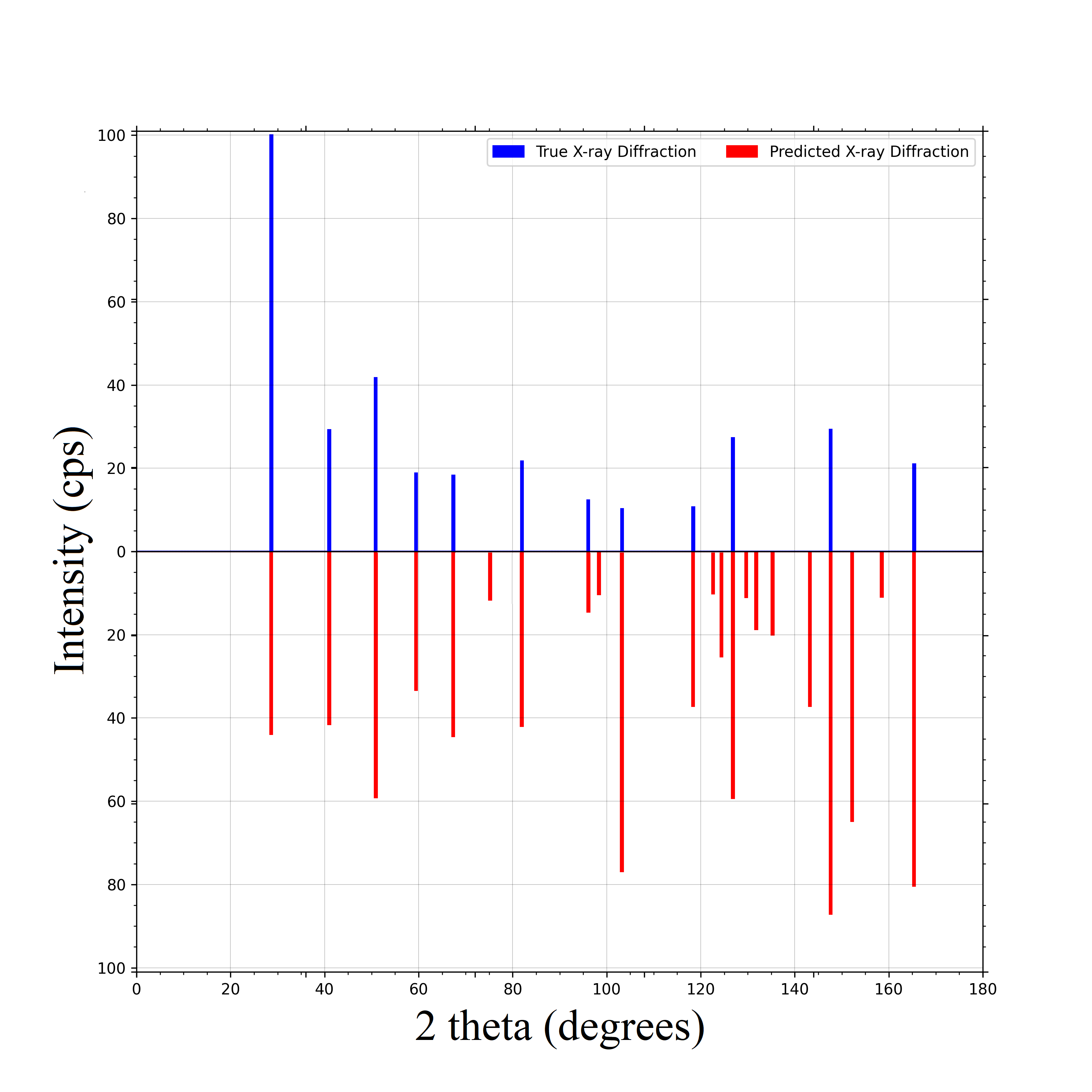}
    \caption{shifted peak XRD VS ground truth}
    \vspace{3pt}
  \end{subfigure} 
  \caption{XRD prediction performance of BaTbO$_3$.}
    \label{fig:prediction performance}
\end{figure}

Figure \ref{fig:prediction performance} shows an example of prediction results of BaTbO$_3$. Figure \ref{fig:prediction performance} (a) shows the predicted XRD values and the target values, we find that our DeepXRD model can find position for most peaks. Figure \ref{fig:prediction performance} (b) only focus on all peak positions in the predicted and target XRD, and we use peak alignment to fine-tune peak positions as Figure \ref{fig:prediction performance} (c) shows. Our algorithm have found almost all XRD peak positions although the peak magnitudes maybe not very accurate.   

\FloatBarrier
\subsection{Hyper-parameter tuning of DeepXRD models}

To obtain the best hyper-parameters for the DeepXRD model, we compare peak position match percentages under different hyper-parameter combinations on two datasets. For a given formula, we first predict its XRD spectrum and then determine all peak positions with magnitude greater than 10. The performance results are shown in Table \ref{table:hyperparameter}. For models with 10, 15, 20, and 25 ResNet layers, we calculate and compare the peak position match percentages with different learning rates and batch sizes. From Table \ref{table:hyperparameter}, we find that for the $ABC_3$-XRD dataset, the model with 20 ResNet layers, learning rate 0.001, and batch size 64 achieves the best performance. After peak shift operations, the final average peak match percentage is 68\%. For the Ternary-XRD dataset (Table \ref{table:hyperparameter_2}), the model with 20 ResNet layers, learning rate 0.001, and batch size 128 achieves the best performance with peak match percentage 63\%.

\begin{table}[H]
\caption{Prediction performance (peak position match percentage) of different parameter settings on the $ABC_3$-XRD dataset.}
\label{table:hyperparameter}
\centering
\resizebox{\textwidth}{!}{\begin{tabular}{c|r|r|r|r|r|r|r|r|r|r|r|r|}
\cline{2-13}
\multicolumn{1}{l|}{} & \multicolumn{3}{c|}{\textbf{\begin{tabular}[c]{@{}c@{}}Learning Rate\\ 0.001\end{tabular}}} & \multicolumn{3}{c|}{\textbf{\begin{tabular}[c]{@{}c@{}}Learning Rate\\ 0.002\end{tabular}}} & \multicolumn{3}{c|}{\textbf{\begin{tabular}[c]{@{}c@{}}Learning Rate\\ 0.003\end{tabular}}} & \multicolumn{3}{c|}{\textbf{\begin{tabular}[c]{@{}c@{}}Learning Rate\\ 0.004\end{tabular}}} \\ \hline
\multicolumn{1}{|c|}{\textbf{\begin{tabular}[c]{@{}c@{}}ResNet\\ Layers\end{tabular}}} &
\multicolumn{1}{c|}{\textit{\begin{tabular}[c]{@{}c@{}}Batch Size \\ 32\end{tabular}}} &
\multicolumn{1}{c|}{\textit{\begin{tabular}[c]{@{}c@{}}Batch Size \\ 64\end{tabular}}} & \multicolumn{1}{c|}{\textit{\begin{tabular}[c]{@{}c@{}}Batch Size\\ 128\end{tabular}}} &
\multicolumn{1}{c|}{\textit{\begin{tabular}[c]{@{}c@{}}Batch Size \\ 32\end{tabular}}} &
\multicolumn{1}{c|}{\textit{\begin{tabular}[c]{@{}c@{}}Batch Size \\ 64\end{tabular}}} & \multicolumn{1}{c|}{\textit{\begin{tabular}[c]{@{}c@{}}Batch Size\\ 128\end{tabular}}} &
\multicolumn{1}{c|}{\textit{\begin{tabular}[c]{@{}c@{}}Batch Size \\ 32\end{tabular}}} &
\multicolumn{1}{c|}{\textit{\begin{tabular}[c]{@{}c@{}}Batch Size\\ 64\end{tabular}}} & \multicolumn{1}{c|}{\textit{\begin{tabular}[c]{@{}c@{}}Batch Size\\ 128\end{tabular}}} &
\multicolumn{1}{c|}{\textit{\begin{tabular}[c]{@{}c@{}}Batch Size \\ 32\end{tabular}}} &
\multicolumn{1}{c|}{\textit{\begin{tabular}[c]{@{}c@{}}Batch Size\\ 64\end{tabular}}} & \multicolumn{1}{c|}{\textit{\begin{tabular}[c]{@{}c@{}}Batch Size \\ 128\end{tabular}}}  \\ \hline
\multicolumn{1}{|c|}{\textbf{10}} & 0.58  & 0.63  & 0.57  & 0.6 & 0.64 & 0.57  & 0.58 & 0.64 & 0.60   &0.59 & 0.61&0.55  \\ \hline
\multicolumn{1}{|c|}{\textbf{15}} &0.58  & 0.63  & 0.63  & 0.57  & 0.62  & 0.59  & 0.58 & 0.63  & 0.65 & 0.56 & 0.59 & 0.53    \\ \hline
\multicolumn{1}{|c|}{\textbf{20}}   & 0.57  & \textbf{0.68}  & 0.59  &  0.58 & 0.65  & 0.58 & 0.56  &0.61  &0.57 & 0.54& 0.59& 0.53  \\ \hline
\multicolumn{1}{|c|}{\textbf{25}}  & 0.54  & 0.61  & 0.59  & 0.57  & 0.59  & 0.59 & 0.59  & 0.59 & 0.56& 0.59 &0.56 &0.53 \\ \hline
\end{tabular}}
\end{table}

\begin{table}[H]
\caption{Prediction performance (peak position match percentage) of different parameter settings on the Ternary-XRD dataset}
\label{table:hyperparameter_2}
\centering
\resizebox{\textwidth}{!}{\begin{tabular}{c|r|r|r|r|r|r|r|r|r|r|r|r|}
\cline{2-13}
\multicolumn{1}{l|}{} & \multicolumn{3}{c|}{\textbf{\begin{tabular}[c]{@{}c@{}}Learning Rate\\ 0.001\end{tabular}}} & \multicolumn{3}{c|}{\textbf{\begin{tabular}[c]{@{}c@{}}Learning Rate\\ 0.002\end{tabular}}} & \multicolumn{3}{c|}{\textbf{\begin{tabular}[c]{@{}c@{}}Learning Rate\\ 0.003\end{tabular}}} & \multicolumn{3}{c|}{\textbf{\begin{tabular}[c]{@{}c@{}}Learning Rate\\ 0.004\end{tabular}}} \\ \hline
\multicolumn{1}{|c|}{\textbf{\begin{tabular}[c]{@{}c@{}}ResNet\\ Layers\end{tabular}}} &
\multicolumn{1}{c|}{\textit{\begin{tabular}[c]{@{}c@{}}Batch Size \\ 64\end{tabular}}} &
\multicolumn{1}{c|}{\textit{\begin{tabular}[c]{@{}c@{}}Batch Size \\ 128\end{tabular}}} & \multicolumn{1}{c|}{\textit{\begin{tabular}[c]{@{}c@{}}Batch Size\\ 256\end{tabular}}} &
\multicolumn{1}{c|}{\textit{\begin{tabular}[c]{@{}c@{}}Batch Size \\ 64\end{tabular}}} &
\multicolumn{1}{c|}{\textit{\begin{tabular}[c]{@{}c@{}}Batch Size \\ 128\end{tabular}}} & \multicolumn{1}{c|}{\textit{\begin{tabular}[c]{@{}c@{}}Batch Size\\ 256\end{tabular}}} &
\multicolumn{1}{c|}{\textit{\begin{tabular}[c]{@{}c@{}}Batch Size \\ 64\end{tabular}}} &
\multicolumn{1}{c|}{\textit{\begin{tabular}[c]{@{}c@{}}Batch Size\\ 128\end{tabular}}} & \multicolumn{1}{c|}{\textit{\begin{tabular}[c]{@{}c@{}}Batch Size\\ 256\end{tabular}}} &
\multicolumn{1}{c|}{\textit{\begin{tabular}[c]{@{}c@{}}Batch Size \\ 64\end{tabular}}} &
\multicolumn{1}{c|}{\textit{\begin{tabular}[c]{@{}c@{}}Batch Size\\ 128\end{tabular}}} & \multicolumn{1}{c|}{\textit{\begin{tabular}[c]{@{}c@{}}Batch Size \\ 256\end{tabular}}}  \\ \hline
\multicolumn{1}{|c|}{\textbf{10}} & 0.57 & 0.57  & 0.57  &  0.57 & 0.57  & 0.57  & 0.57  & 0.59 & 0.57  & 0.55  &0.56  &0.57 \\ \hline
\multicolumn{1}{|c|}{\textbf{15}} & 0.57  & 0.59  & 0.57  & 0.57  &  0.57 & 0.57  & 0.57  & 0.58 & 0.58  & 0.56  & 0.56  & 0.54\\ \hline
\multicolumn{1}{|c|}{\textbf{20}} & 0.61  & \textbf{0.63}  & 0.6  &  0.57 & 0.58  & 0.58  & 0.57  & 0.59 & 0.58  & 0.53  & 0.58  & 0.57 \\ \hline
\multicolumn{1}{|c|}{\textbf{25}} & 0.6  & 0.61  & 0.59  &  0.53 & 0.59  & 0.56  & 0.52  & 0.58 & 0.51  & 0.56  & 0.59  & 0.57\\ \hline
\end{tabular}}
\end{table}

\subsection{Case studies of DeepXRD for XRD spectrum prediction}

\begin{figure}[hb!] 
    \begin{subfigure}[t]{0.28\textwidth}
        \includegraphics[width=\textwidth]{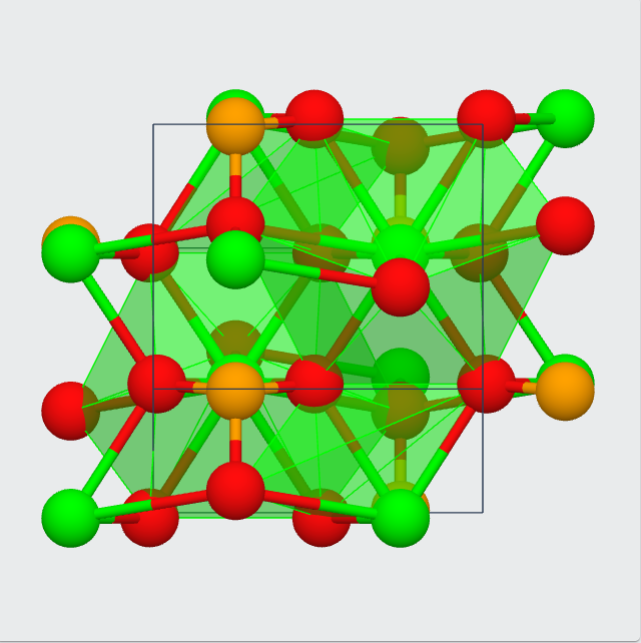}
        \caption{SrSeO$_3$}
        \vspace{-3pt}
    \end{subfigure}\hfill
    \begin{subfigure}[t]{0.28\textwidth}
        \includegraphics[width=\textwidth]{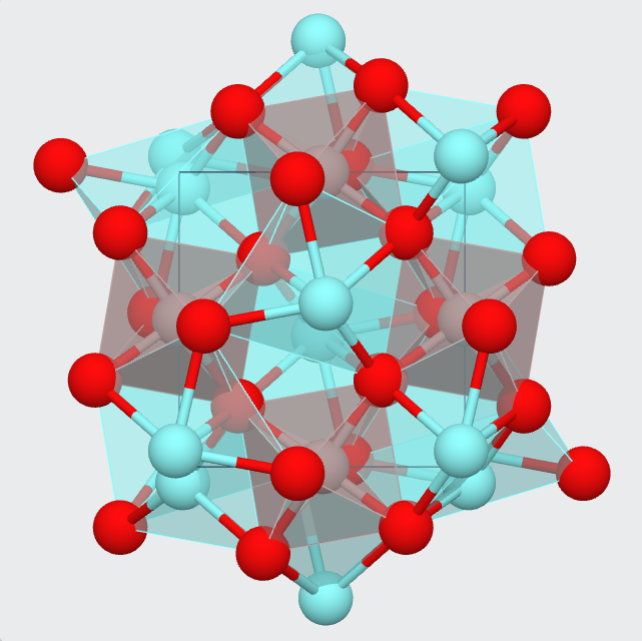}
        \caption{YAlO$_3$}
        \vspace{-3pt}
    \end{subfigure}\hfill
    \begin{subfigure}[t]{0.28\textwidth}
        \includegraphics[width=\textwidth]{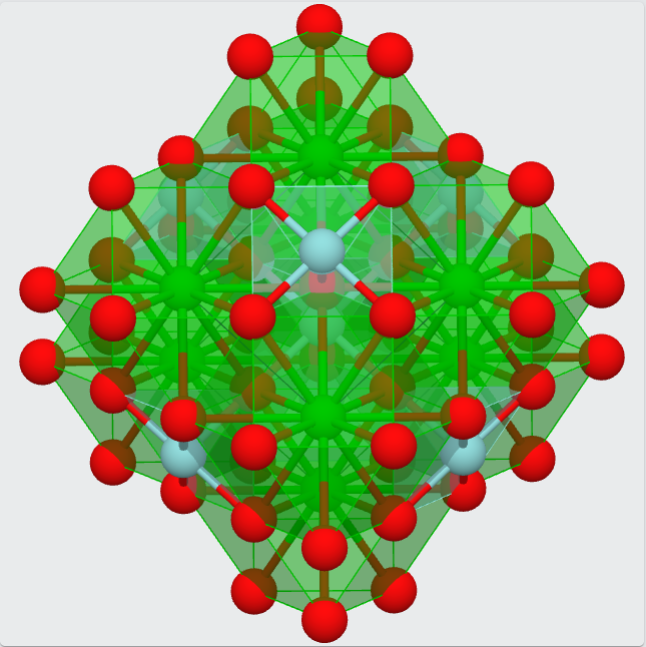} 
        \caption{BaZrO$_3$}
        \vspace{-3pt}
    \end{subfigure}\hfill
    \vspace{1pt}
    \begin{subfigure}[t]{0.28\textwidth}
        \includegraphics[width=\textwidth]{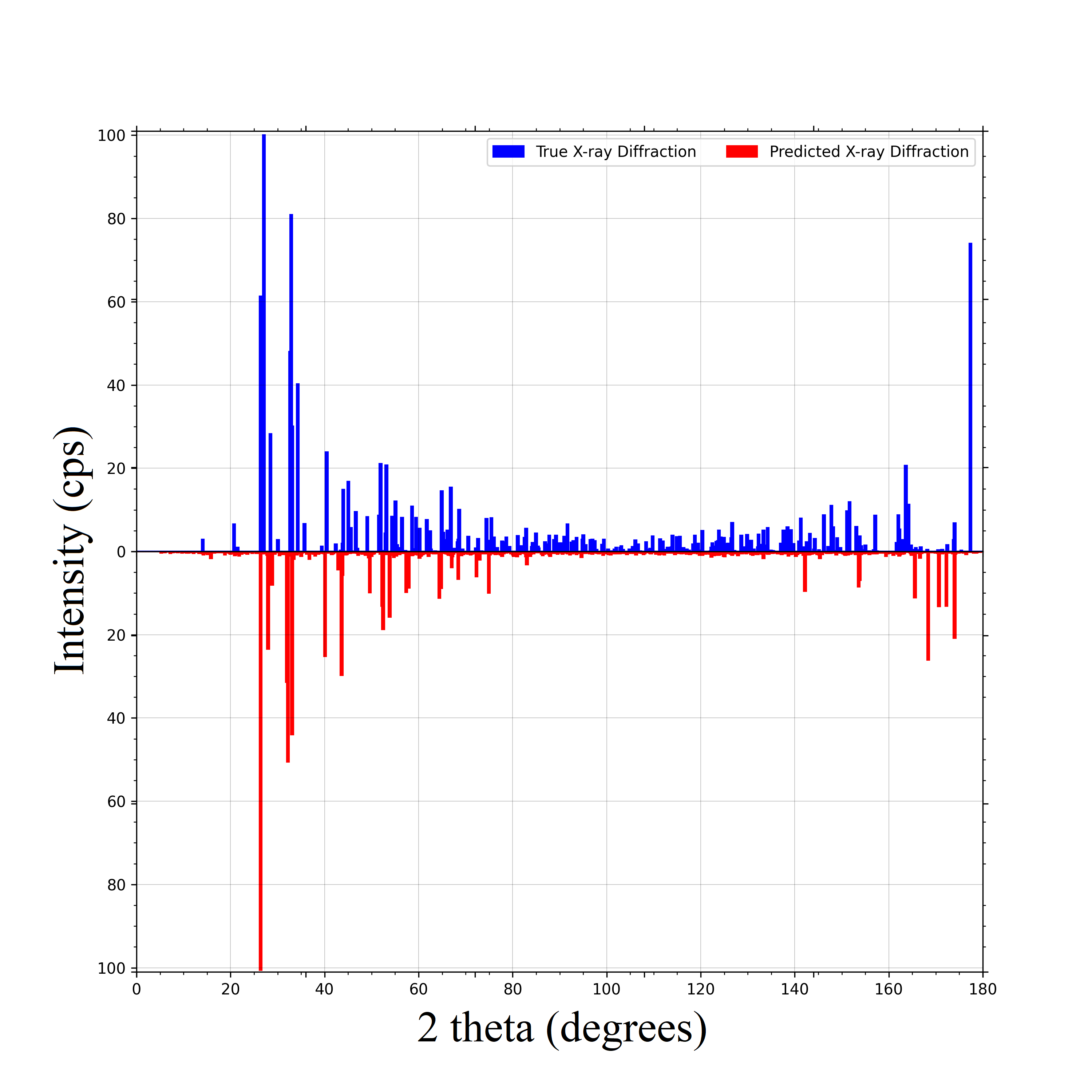}
        \caption{SrSeO$_3$ predicted XRD}
        \vspace{-3pt}
    \end{subfigure}\hfill    
    \begin{subfigure}[t]{0.28\textwidth}
        \includegraphics[width=\textwidth]{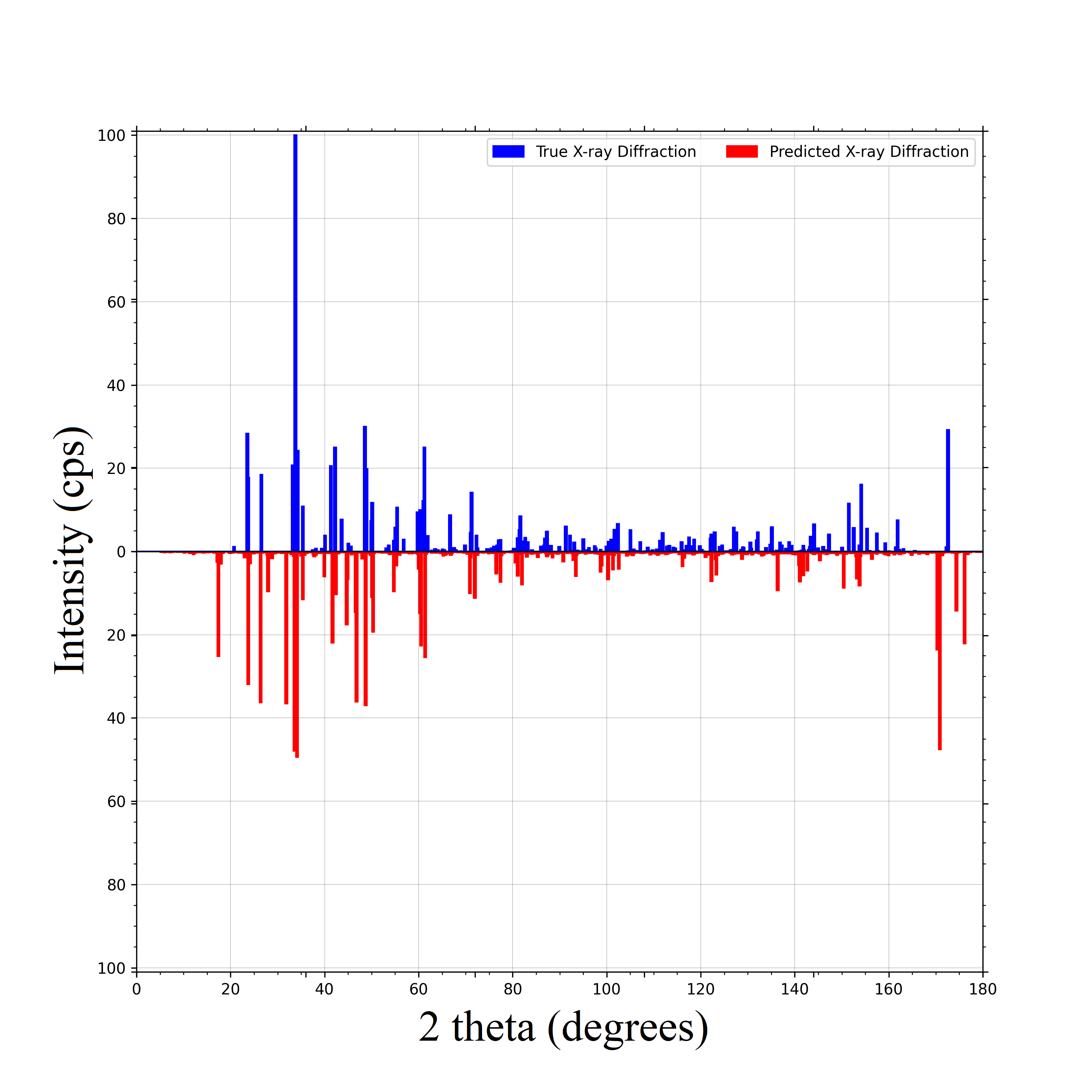}
        \caption{YAlO$_3$ predicted XRD}
        \vspace{-3pt}
        \label{fig:Ce2As2O6_predim}
    \end{subfigure}\hfill
    \begin{subfigure}[t]{0.28\textwidth}
        \includegraphics[width=\textwidth]{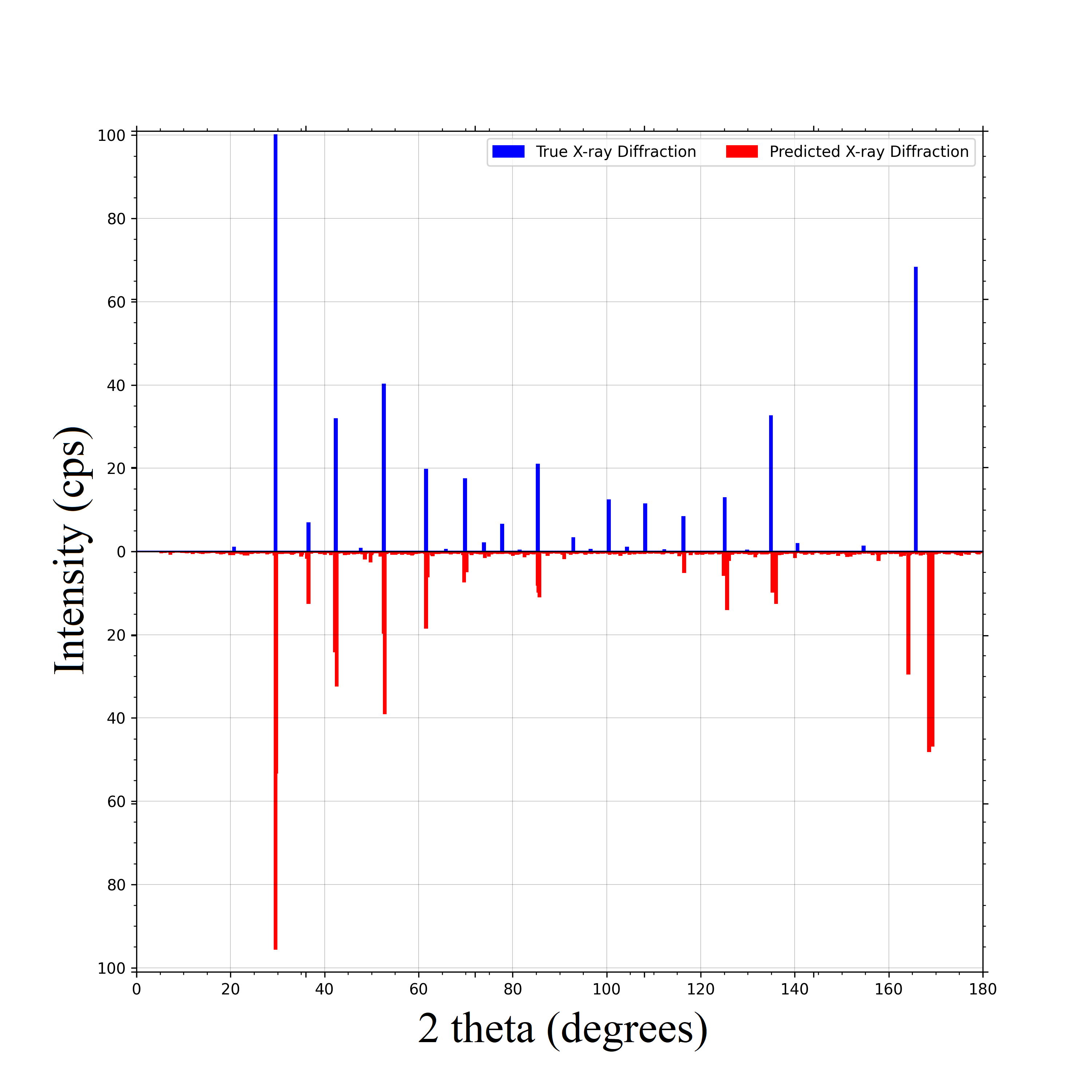}
        \caption{BaZrO$_3$ predicted XRD}
        \vspace{-3pt}
    \end{subfigure}\hfill
    \begin{subfigure}[t]{0.28\textwidth}
        \includegraphics[width=\textwidth]{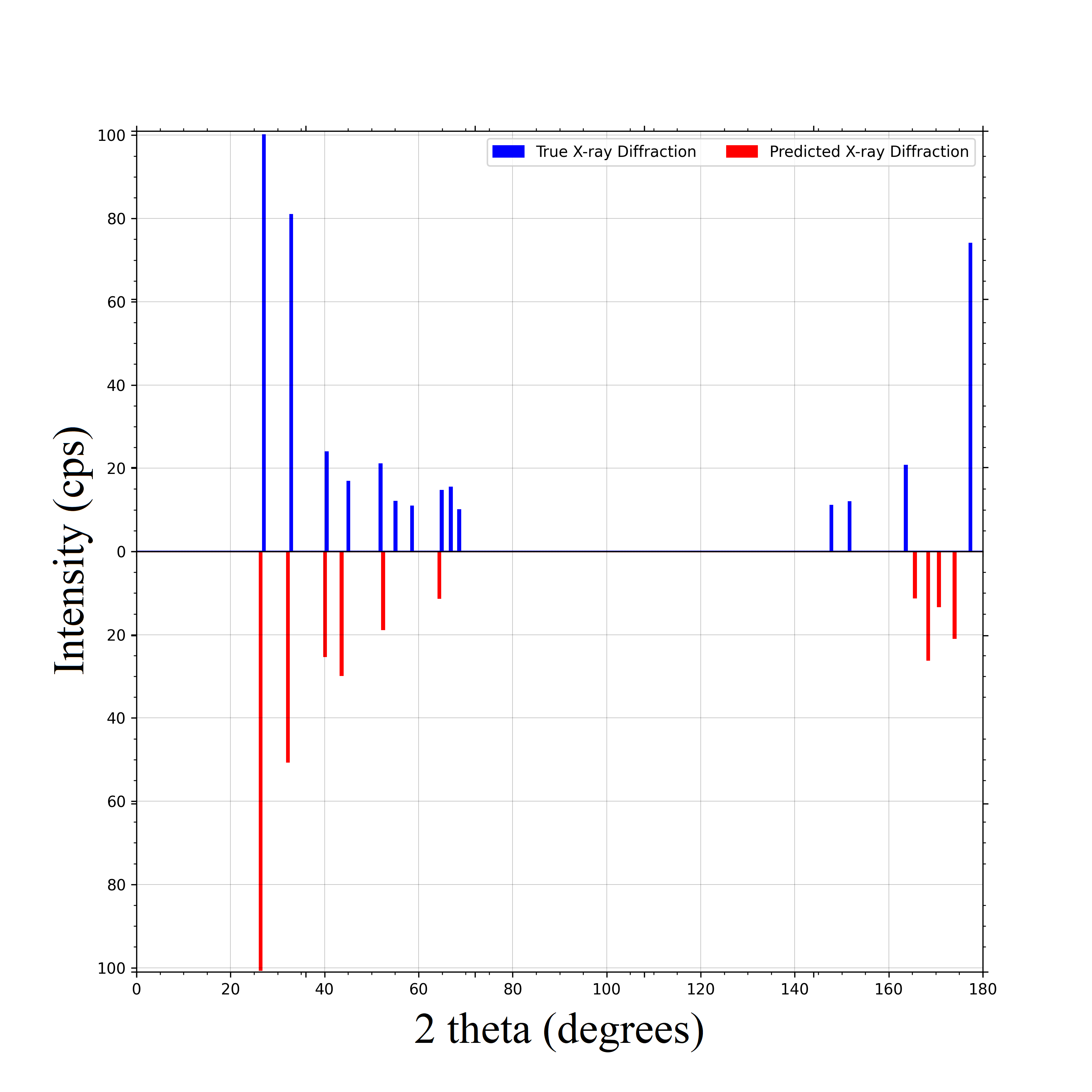}
        \caption{SrSeO$_3$ predicted XRD peak}
        \vspace{-3pt}
    \end{subfigure}\hfill
    \begin{subfigure}[t]{0.28\textwidth}
        \includegraphics[width=\textwidth]{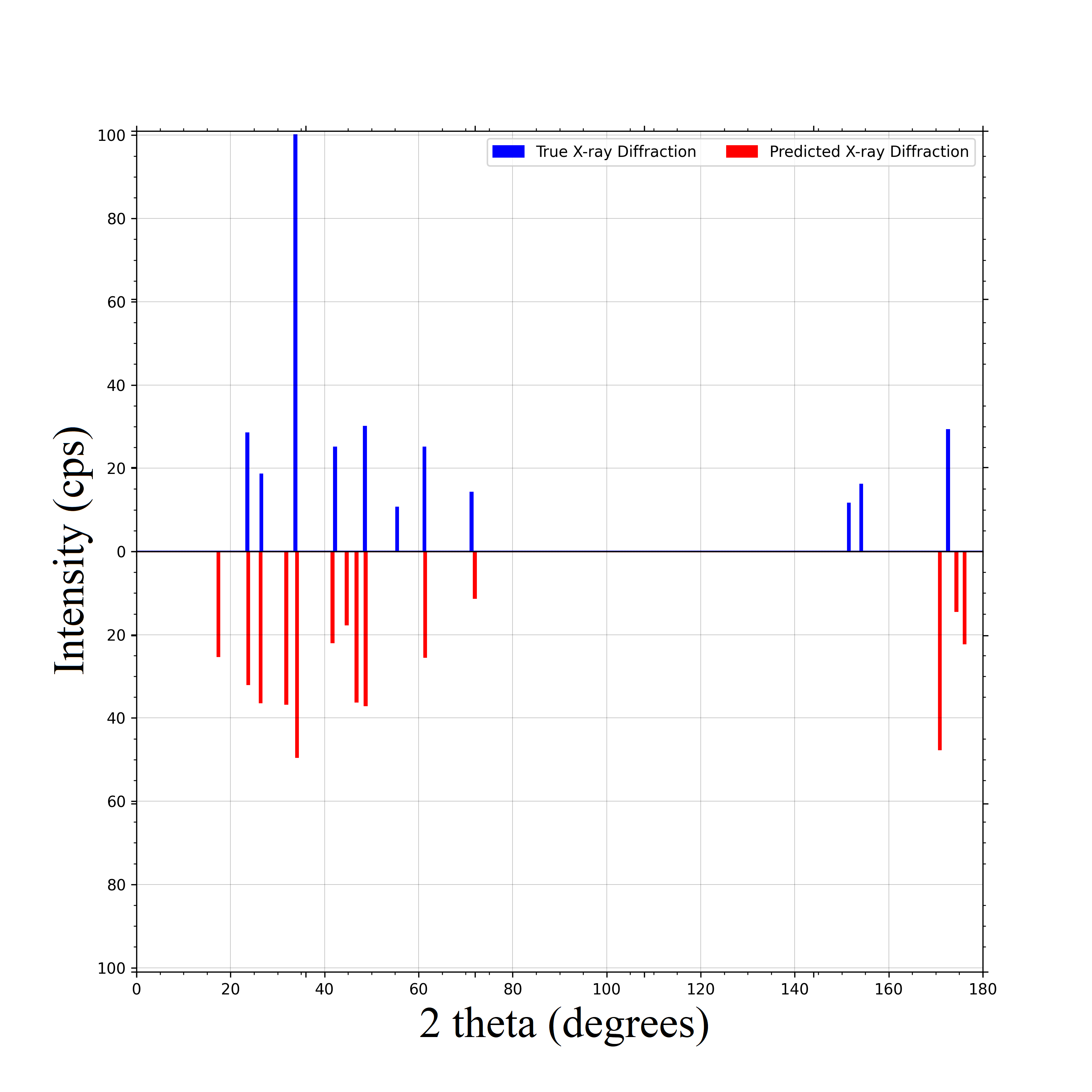}
        \caption{YAlO$_3$ predicted XRD peak}
        \vspace{-3pt}
        \label{fig:k}
    \end{subfigure}\hfill
    \begin{subfigure}[t]{0.28\textwidth}
        \includegraphics[width=\textwidth]{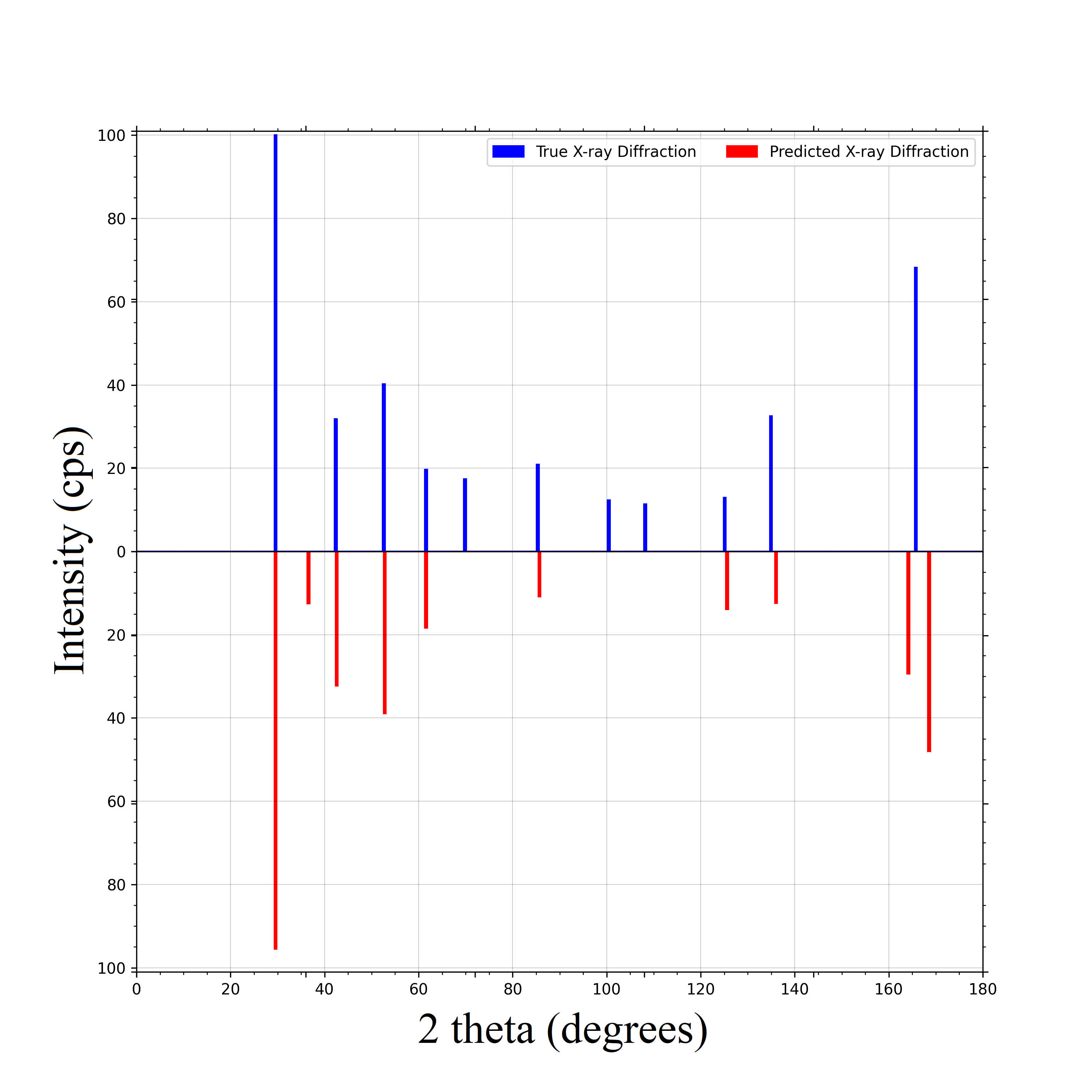}
        \caption{BaZrO$_3$ predicted XRD peak}
        \vspace{-3pt}
        \label{fig:l}
    \end{subfigure}\hfill
    \begin{subfigure}[t]{0.28\textwidth}
        \includegraphics[width=\textwidth]{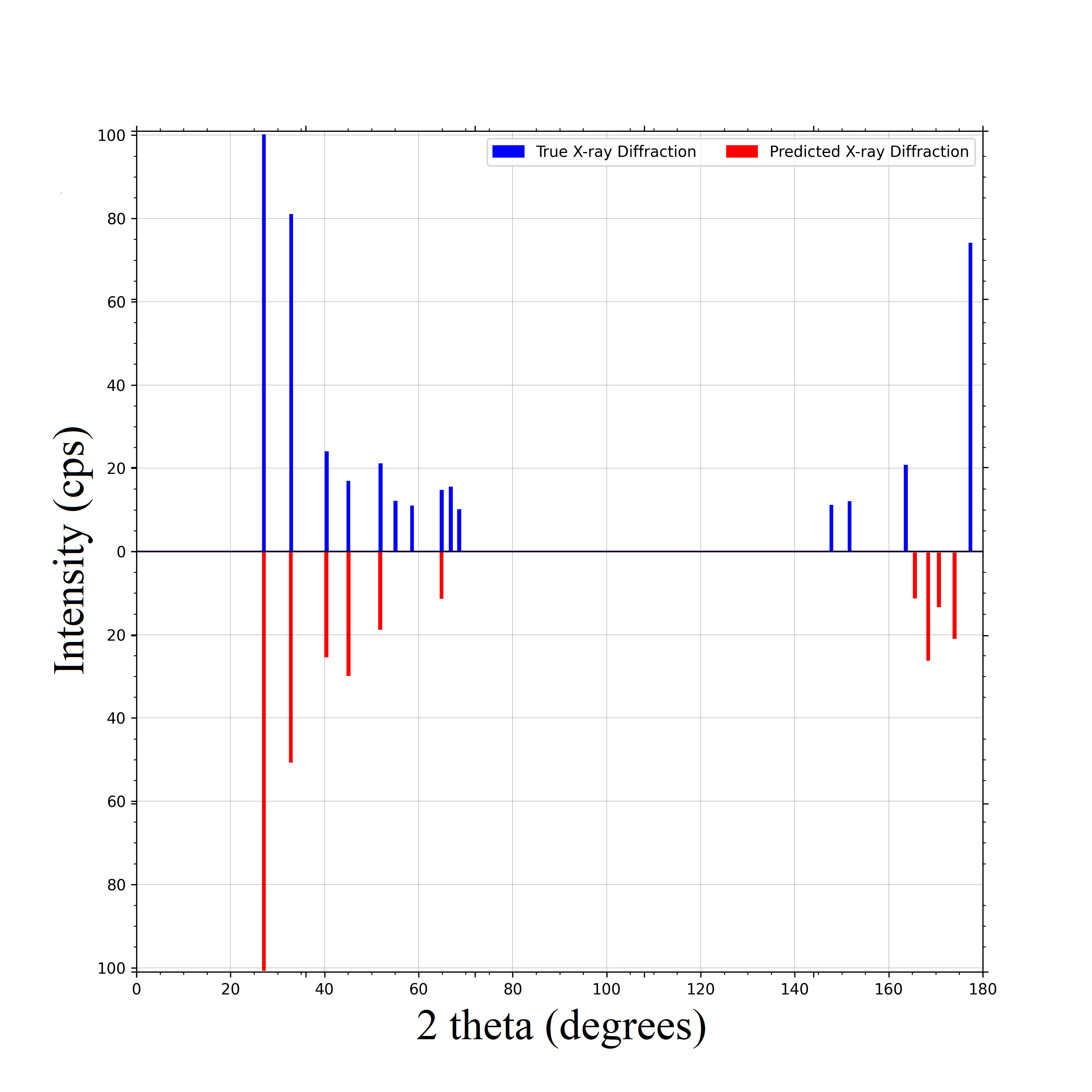}
        \caption{BaSnO$_3$ alignment XRD peak}
        \vspace{-3pt}
        \label{fig:g}
    \end{subfigure}\hfill    
    \begin{subfigure}[t]{0.28\textwidth}
        \includegraphics[width=\textwidth]{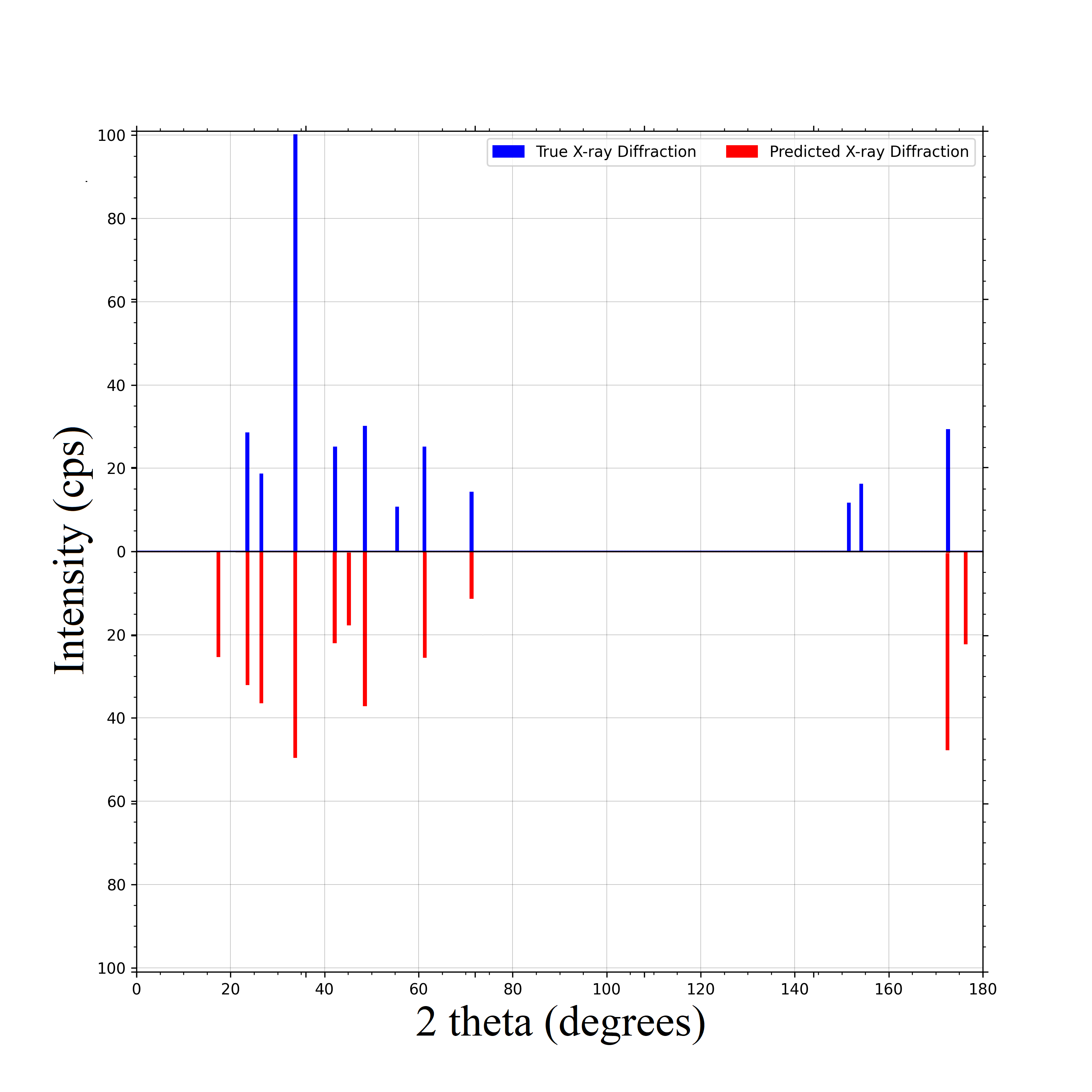}
        \caption{YAlO$_3$ alignment XRD peak}
        \vspace{-3pt}
        \label{fig:h}
    \end{subfigure}\hfill
    \begin{subfigure}[t]{0.28\textwidth}
        \includegraphics[width=\textwidth]{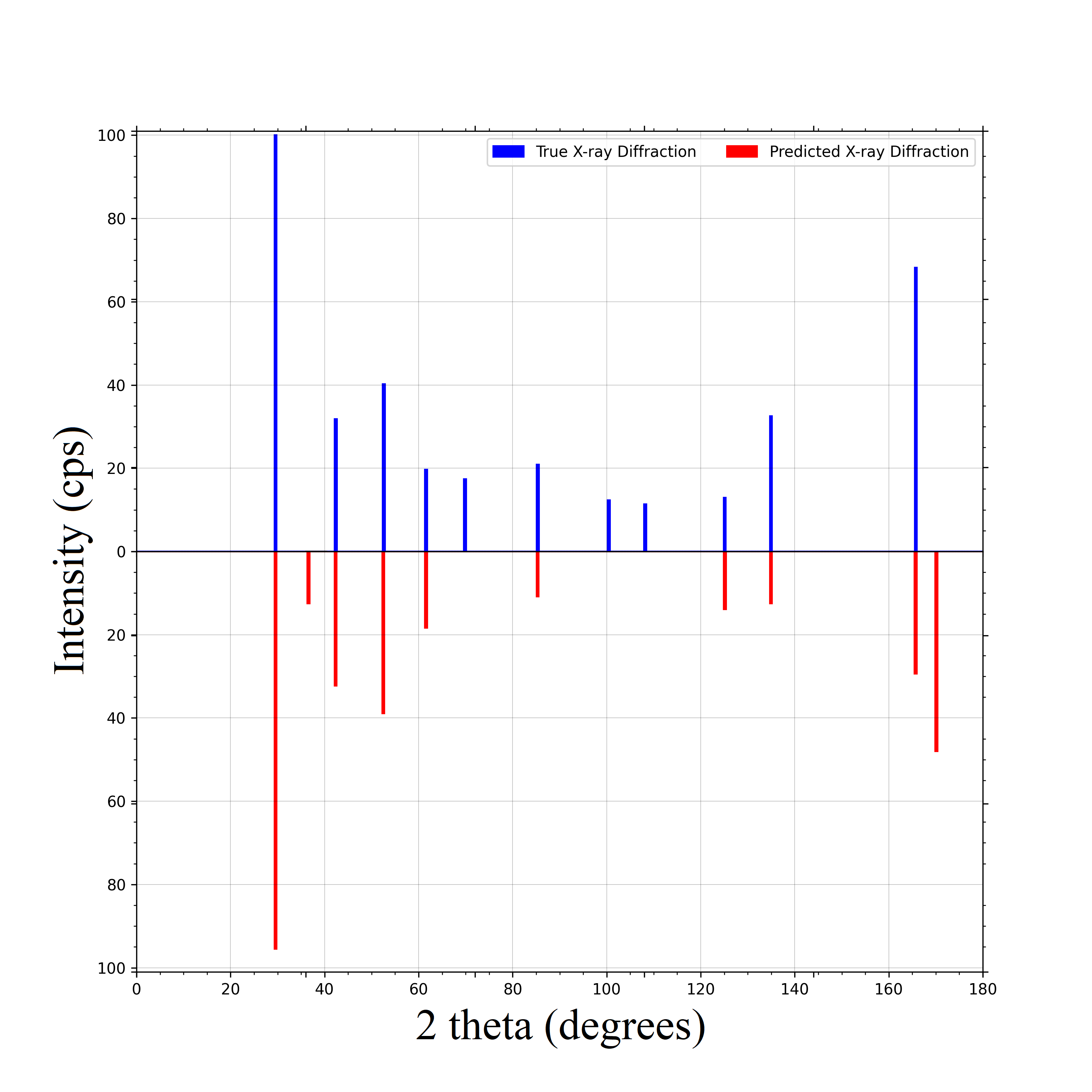}
        \caption{BaZrO$_3$ alignment XRD peak}
        \vspace{-3pt}
        \label{fig:i}
    \end{subfigure}\hfill   
    \caption{Prediction performance of peak positions by DeepXRD. (a)(d)(g)(j): structure and predicted peak positions of BaSnO$_3$; (b)(e)(h)(k):structure and predicted peak positions of YAlO$_3$; (c)(f)(i)(l): structure and peak positions of BaZrO$_3$.}
    \label{fig:case_studies_performance}
\end{figure}

To evaluate the performance of our DeepXRD model, we randomly select three target compositions and their XRDs as design targets from the $ABC_3$-XRD test set. The XRDs of the formulas predicted by the DeepXRD model (red color) are shown in Figure \ref{fig:case_studies_performance} together with the target XRDs (blue color). The first row shows the structure of given formulas. The second row shows   XRD values predicted by the DeepXRD model. The third row shows the results that ignore the noise and only focus on peak positions and peak values. The last row shows predicted and ground truth XRD peak positions and peak values after peak alignment.

As most $ABC_3$ structures are ABO\textsubscript{3}, we fix the C in $ABC_3$ as oxygen and choose different A and B to evaluate how DeepXRD models perform on different materials. In the 3D space of crystal materials, there are seven crystal systems: triclinic, monoclinic, orthorhombic, tetragonal, trigonal, hexagonal, and cubic. We choose three samples from the most common crystal systems of ABO3: cubic, orthorhombic, and monoclinic respectively to evaluate the predictive performance of our model. The crystal system of SrSeO$_3$ is monoclinic, and the structure of SrSeO$_3$ is shown in Figure \ref{fig:case_studies_performance} (a). It has two high peaks: one is between 25 and 35 degrees; another is around 175 to 180 degrees, and also some small peaks and noises, as shown in Figure \ref{fig:case_studies_performance} (d). The predicted peaks (red color) match well with the approximate positions of the first two high peaks along with some small peaks, and the magnitude of the highest peak is accurately predicted. 
The last high peak located between 175 and 180 degrees is missed, which is probably due to we do not have many training samples containing elements Sr and Se that generate peaks in this area. That is why our model does not predict a peak at around 180 degrees. After filtering the noise signals in the XRD values (values smaller than 10cps) we obtain Figure \ref{fig:case_studies_performance} (g), which shows the peak matches much better than (d) by focusing only on significant peaks. From this figure, we find that there are small gaps in terms of the peak positions, which is related to the common X-ray diffraction peak shifting phenomena in crystallography \cite{shiah1973kinematic}. To consider this factor, we conduct a peak shifting operation to match and adjust the predicted peaks within a distance smaller than 2 degrees to the ground truth peaks, which makes our predictions much closer to the true values, as shown in Figure \ref{fig:case_studies_performance} (j).

The crystal system of YAlO$_3$ is orthorhombic, and its structure is shown in Figure \ref{fig:case_studies_performance} (b). It has only one high peak, which is around 30 to 35 degrees along with several small peaks. As shown in Figure \ref{fig:case_studies_performance} (e), the predicted peaks (red color) successfully match the approximate positions of almost all peaks with a value greater than 20. After ignoring the noises in the XRD spectrum (values smaller than 10), Figure \ref{fig:case_studies_performance} (h) show our predicted peaks and their matches with the ground truth more clearly. Figure \ref{fig:case_studies_performance} (k) shows the matches after peak alignment: our algorithm predicts the same number of peaks as the true ones and only the smallest three of the eleven peaks are not aligned.
Figure \ref{fig:case_studies_performance} (c) shows the structure of BaZrO$_3$, which is a cubic material with scattered peaks and two high peaks around 30 and 165 degrees respectively. It also has several median peaks and almost no noise. As shown in Figure \ref{fig:case_studies_performance} (f), the predicted result (red color) successfully match the approximate positions and magnitudes of the highest peak and almost all peak positions within the first 90 degrees. Ignoring the noises in the XRD values (values smaller than 10cps) does not improve the prediction results very much as shown in Figure \ref{fig:case_studies_performance} (i). Peak alignment can help to adjust the predicted peak positions in the second half as shown in Figure \ref{fig:case_studies_performance} (l). Our algorithm only misses the smallest three peaks.

From Figure\ref{fig:case_studies_performance}, we can find that the XRD distributions predicted by our DeepXRD model are very similar to true XRD spectrum. When we focus only on the peak positions, the prediction errors can be further reduced. The  peak position matches with the ground truths can be fine-tuned with the peak alignment operations. We also find that if the material crystal system or composition elements of a test material composition are infrequent in the training set, the predicted accuracy may become lower.

\begin{figure}[hb!] 
    \begin{subfigure}[t]{0.28\textwidth}
        \includegraphics[width=\textwidth]{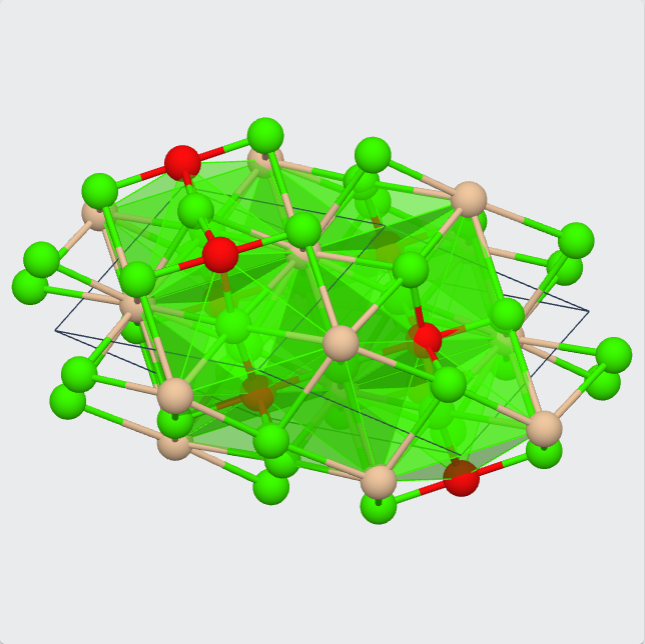}
        \caption{Ca$_3$SiO}
        \vspace{-3pt}
    \end{subfigure}\hfill
    \begin{subfigure}[t]{0.28\textwidth}
        \includegraphics[width=\textwidth]{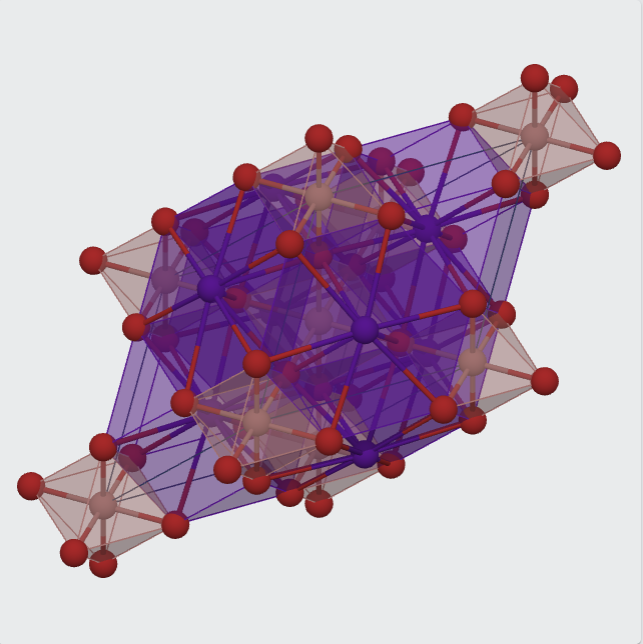}
        \caption{CsInBr$_3$}
        \vspace{-3pt}
    \end{subfigure}\hfill
    \begin{subfigure}[t]{0.28\textwidth}
        \includegraphics[width=\textwidth]{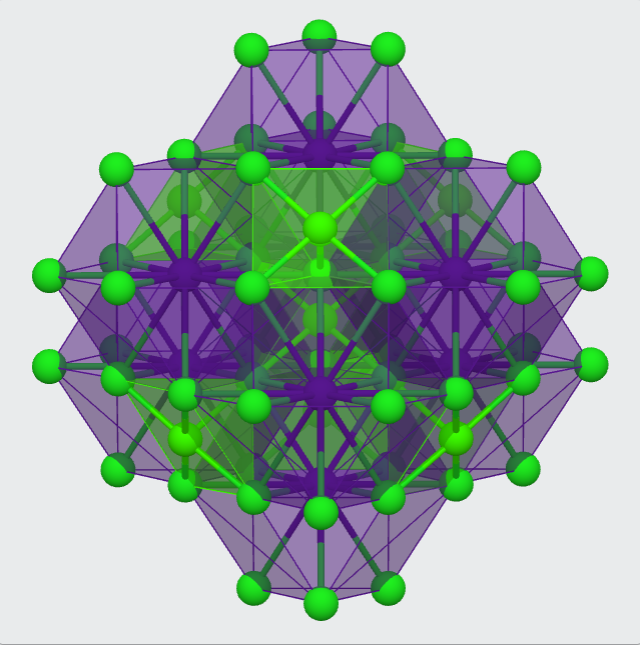} 
        \caption{CsCaCl$_3$}
        \vspace{-3pt}
    \end{subfigure}\hfill
    \vspace{1pt}
    \begin{subfigure}[t]{0.29\textwidth}
        \includegraphics[width=\textwidth]{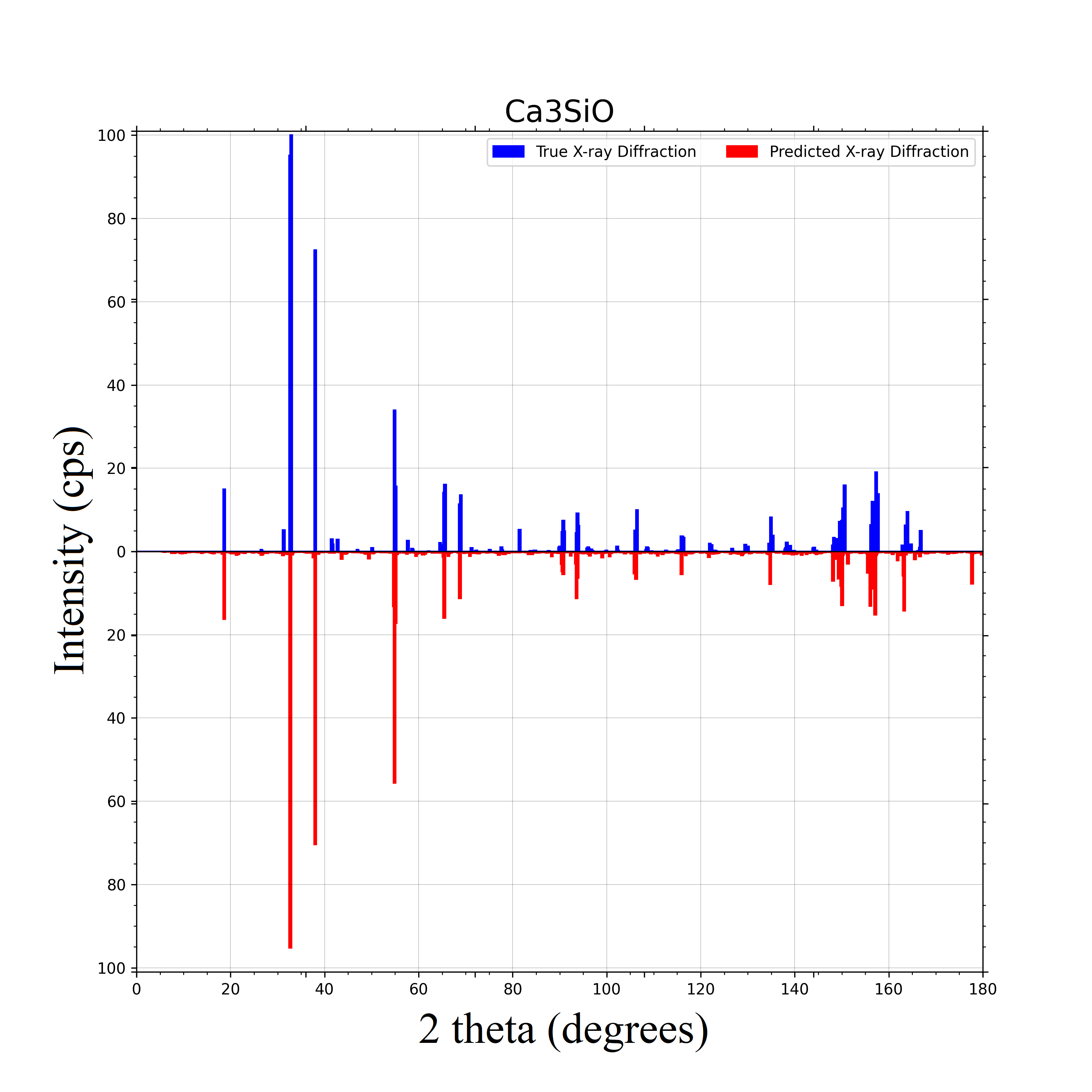}
        \caption{Ca$_3$SiO predicted XRD}
        \vspace{-3pt}
    \end{subfigure}\hfill    
    \begin{subfigure}[t]{0.29\textwidth}
        \includegraphics[width=\textwidth]{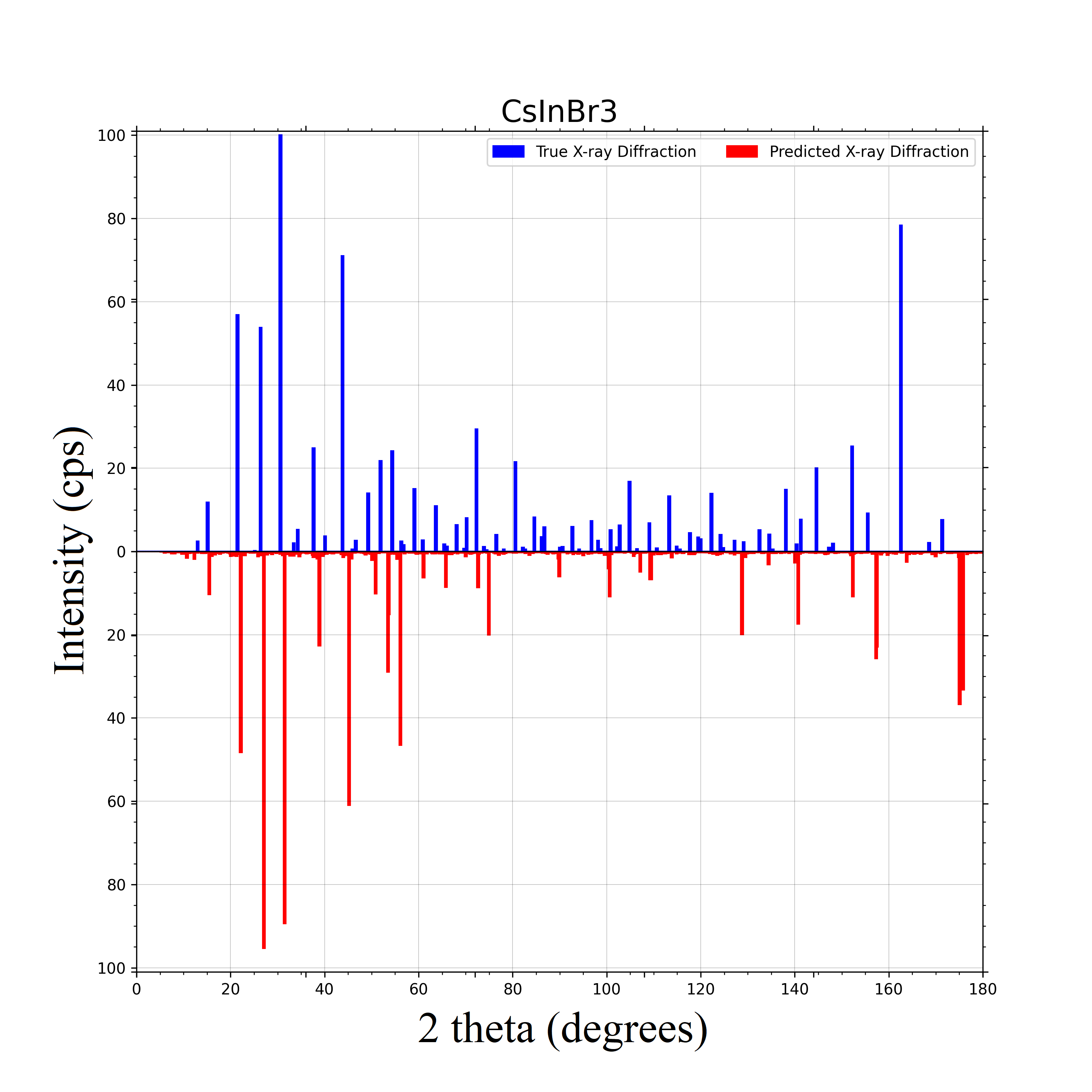}
        \caption{CsInBr$_3$ predicted XRD}
        \vspace{-3pt}
    \end{subfigure}\hfill
    \begin{subfigure}[t]{0.29\textwidth}
        \includegraphics[width=\textwidth]{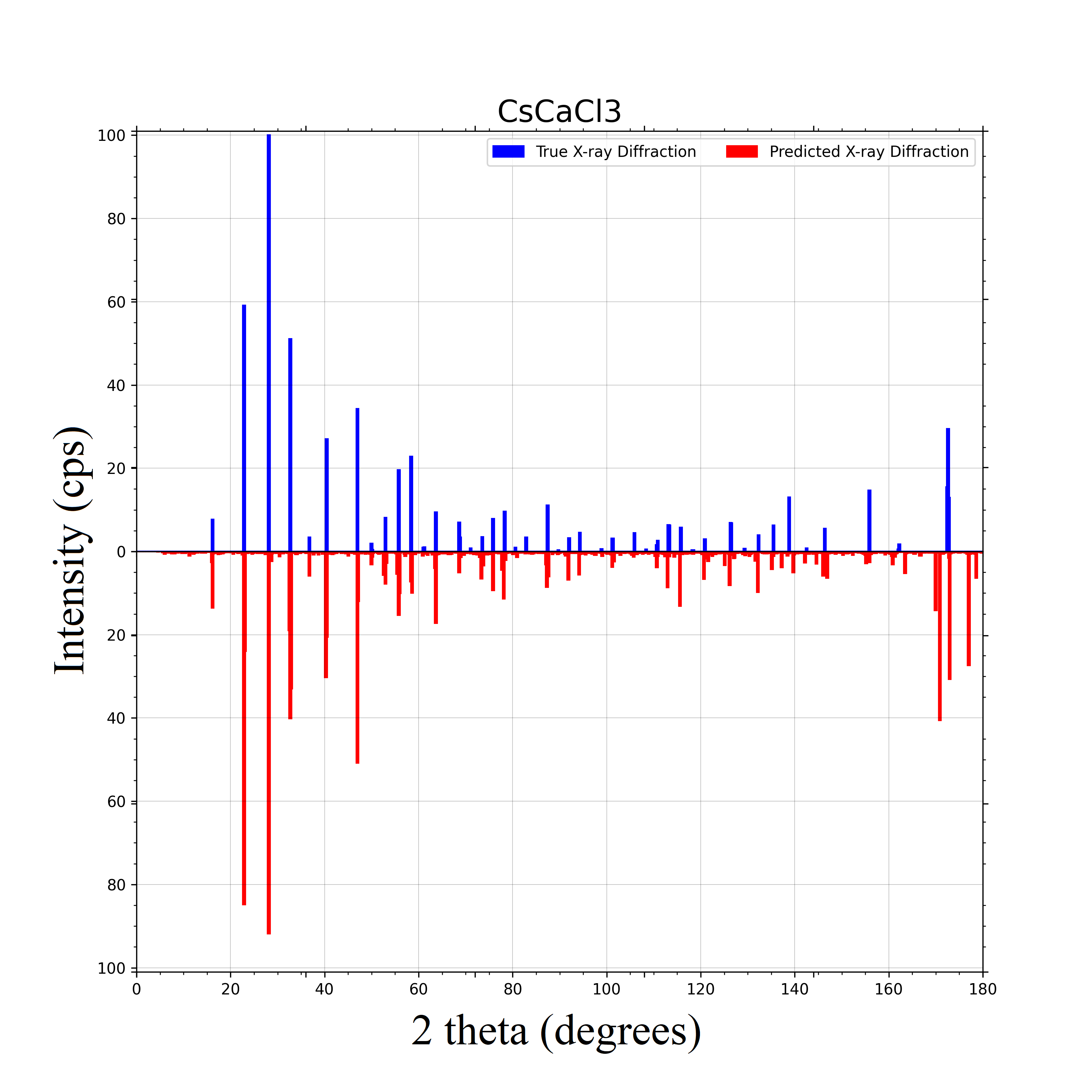}
        \caption{CsCaCl$_3$ predicted XRD}
        \vspace{-3pt}
    \end{subfigure}\hfill
    \begin{subfigure}[t]{0.28\textwidth}
        \includegraphics[width=\textwidth]{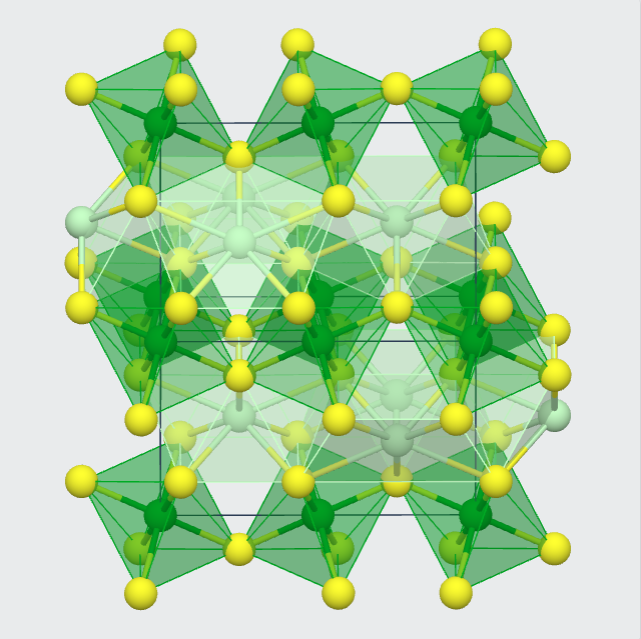}
        \caption{NdLuS$_3$}
        \vspace{-3pt}
    \end{subfigure}\hfill
    \begin{subfigure}[t]{0.28\textwidth}
        \includegraphics[width=\textwidth]{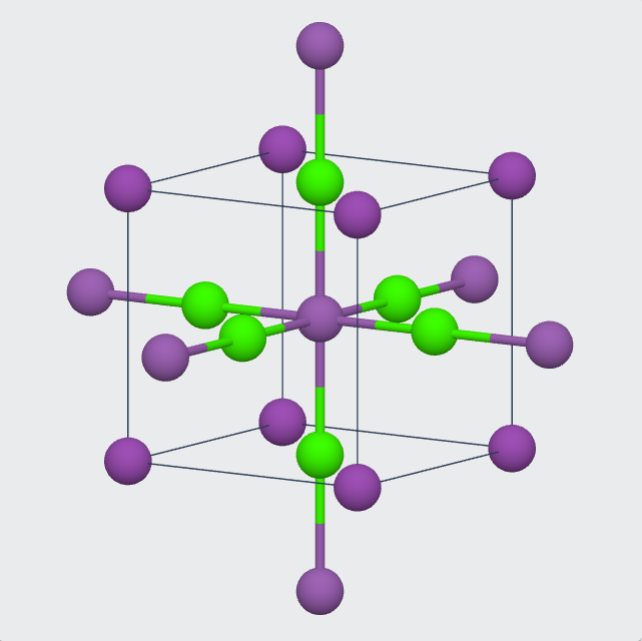}
        \caption{Ca$_3$BiSb}
        \vspace{-3pt}
    \end{subfigure}\hfill
    \begin{subfigure}[t]{0.28\textwidth}
        \includegraphics[width=\textwidth]{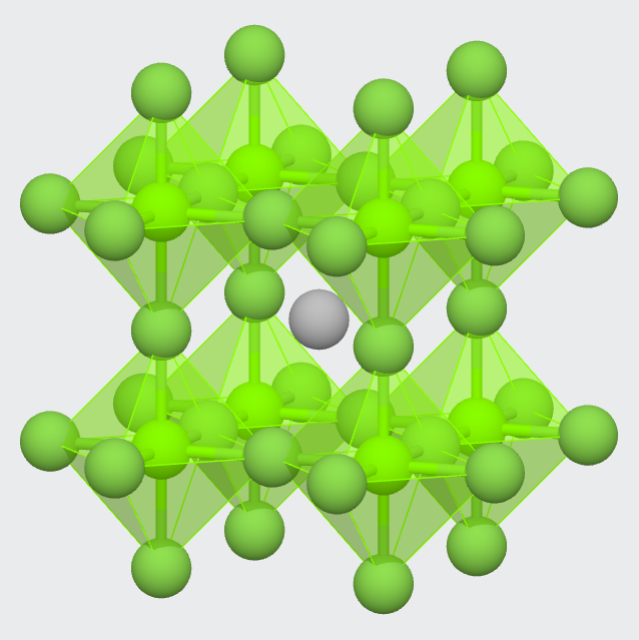}
        \caption{MgAgF$_3$}
        \vspace{-3pt}
    \end{subfigure}\hfill
    \begin{subfigure}[t]{0.29\textwidth}
        \includegraphics[width=\textwidth]{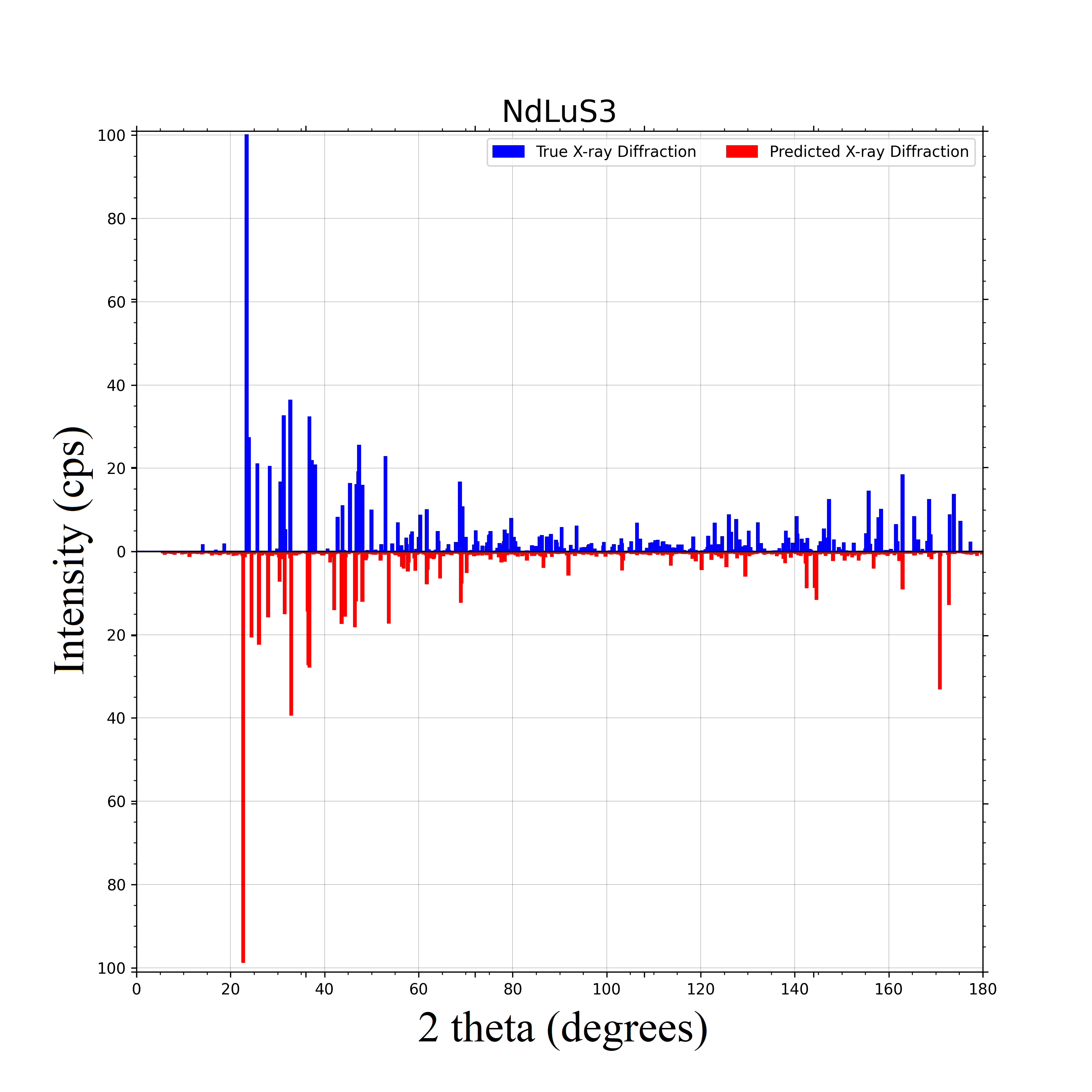}
        \caption{NdLuS$_3$ predicted XRD}
        \vspace{-3pt}
    \end{subfigure}\hfill    
    \begin{subfigure}[t]{0.29\textwidth}
        \includegraphics[width=\textwidth]{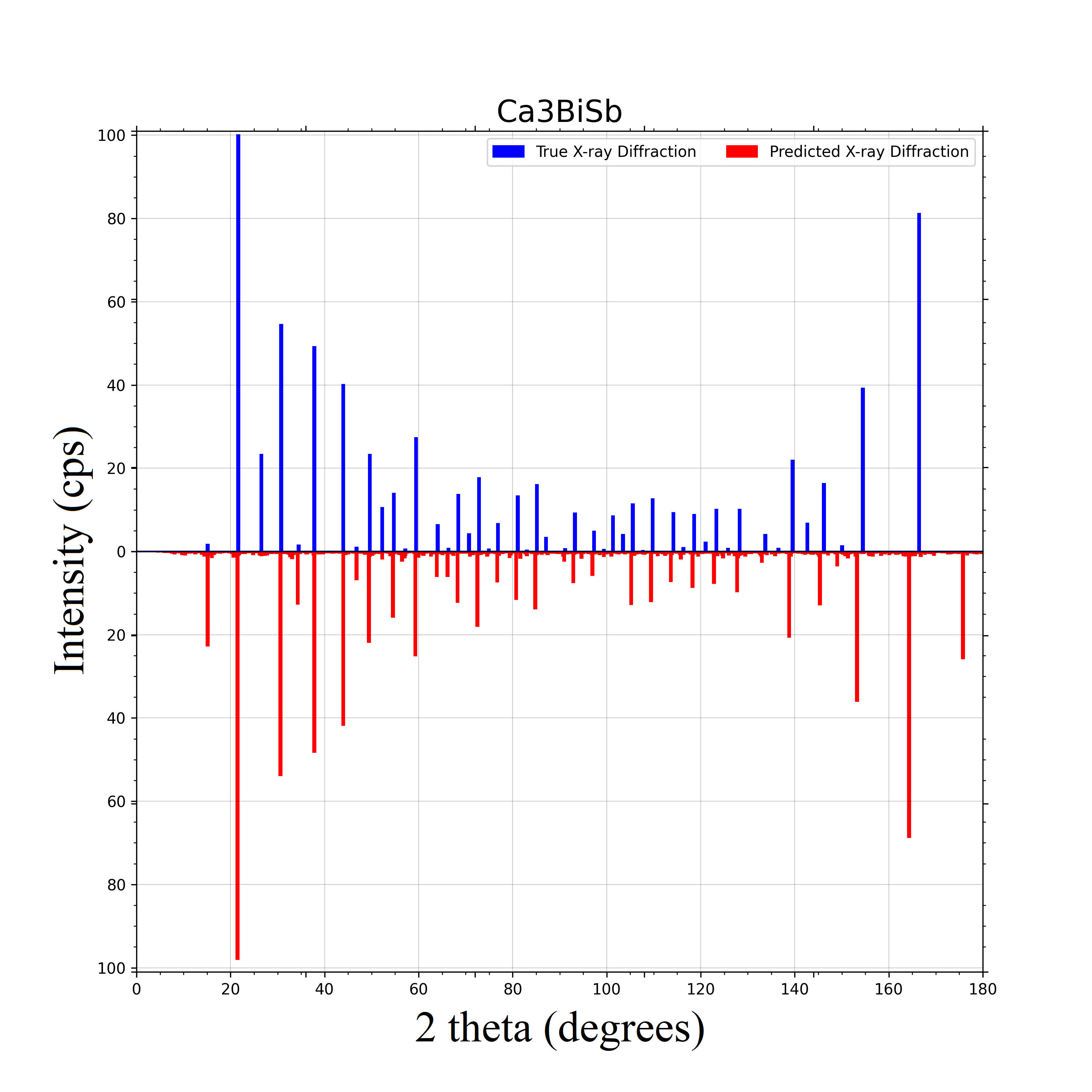}
        \caption{Ca$_3$BiSb predicted XRD}
        \vspace{-3pt}
    \end{subfigure}\hfill
    \begin{subfigure}[t]{0.29\textwidth}
        \includegraphics[width=\textwidth]{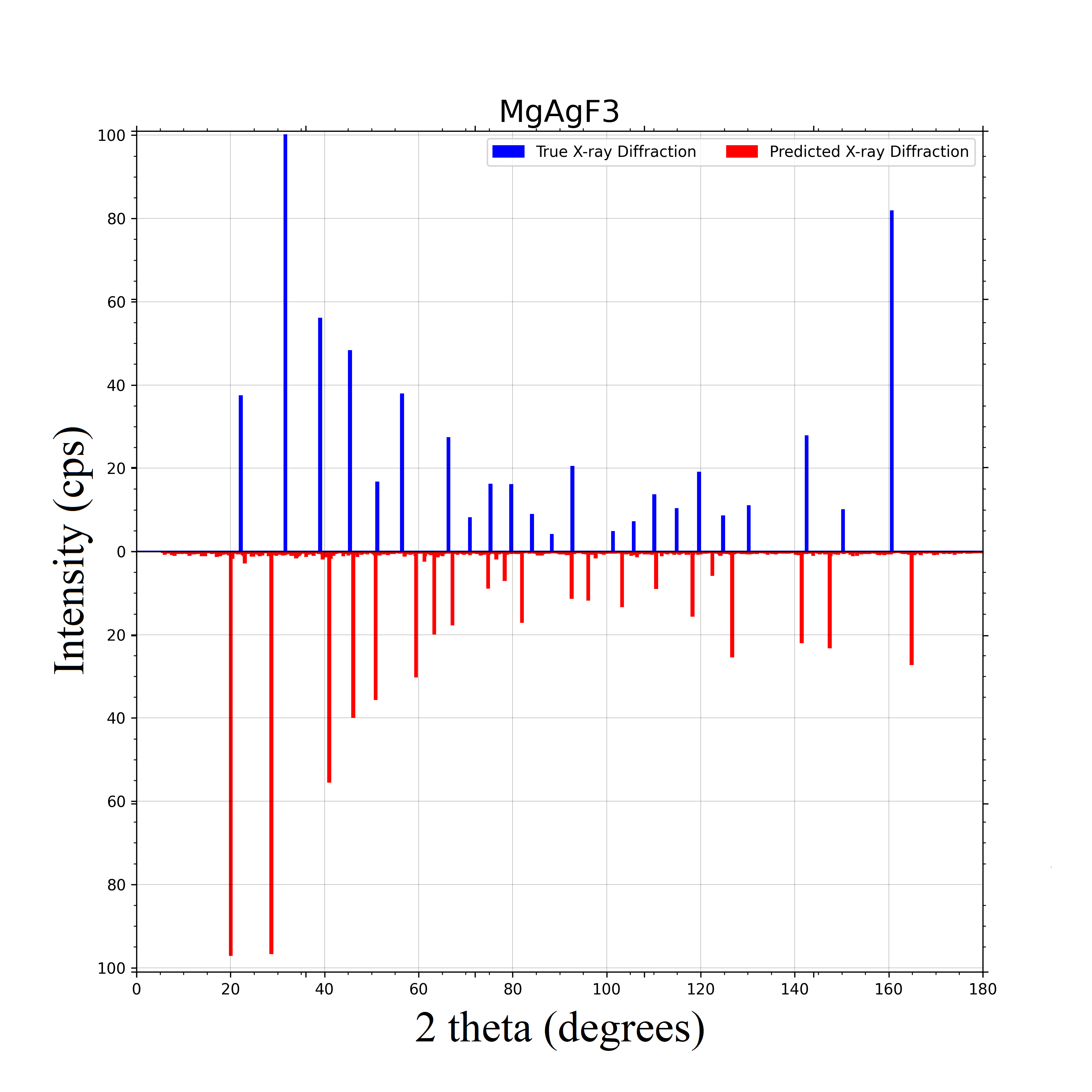}
        \caption{MgAgF$_3$ predicted XRD}
        \vspace{-3pt}
    \end{subfigure}\hfill   
    \caption{Prediction performance of DeepXRD. (a)(d): structure and predicted XRD of Ca$_3$SiO; (b)(e): structure and predicted XRD of CsInBr$_3$; (c)(f): structure and predicted XRD of CsCaCl$_3$; (g)(j): structure and predicted XRD of NdLuS$_3$; (h)(k): structure and predicted XRD of Ca$_3$BiSb; (i)(l): structure and predicted XRD of MgAgF$_3$.}
    \label{fig:case_studies_performance_1}
\end{figure}

To further evaluate the DeepXRD model's performance on the $ABC_3$-XRD dataset, we choose test samples with different A, B and C elements.  Figure \ref{fig:case_studies_performance_1} shows their predicted XRD (red color) and ground truth XRD spectra (blue color). 
The first test sample Ca$_3$SiO is orthorhombic (Figure \ref{fig:case_studies_performance_1} (a)), which has two high-intensity peaks within the interval of 30 to 40 degrees and a median peak around 50 degrees together with several small peaks. Our predicted XRD spectrum matches almost all peaks of Ca$_3$SiO as shown in Figure \ref{fig:case_studies_performance_1} (d). 
The second test case is CsInBr$_3$, which has a cubic structure as shown in Figure \ref{fig:case_studies_performance_1} (b). CsInBr$_3$ has 5 high peaks: the first four peak positions are around 20 to 50 degrees and the last peak is located at 160 to 165 degrees. 
Figure \ref{fig:case_studies_performance_1} (e) shows that for peaks with magnitude values greater than 20cps, our predicted peak intensity and positions are similar to the true ones except the last high peak. 
The third test sample CsCaCl$_3$, as shown in Figure \ref{fig:case_studies_performance_1} (c), is tetragonal. Figure \ref{fig:case_studies_performance_1} (f) shows that CsCaCl$_3$'s main peak is around 20 to 60 degrees. And our model accurately predicted the exactly positions of these peaks. The fourth test sample NdLuS$_3$ is orthorhombic, and its structure is shown in Figure \ref{fig:case_studies_performance_1} (g). Its highest peak is the first peak located around 20 degrees and the remaining peaks are around 20 to 60 degrees as shown in Figure \ref{fig:case_studies_performance_1} (j). Our predicted XRD spectrum matches almost all peak positions with only several peak differences in intensities. Figure \ref{fig:case_studies_performance_1} (h) shows the fifth test case Ca$_3$BiSb, which is a \hl{sparse} cubic with two high peaks: the first and the last one of all peaks with intensity greater than 80cps. Our predicted XRD matches the first and highest peak very well and matches the second highest peak with only small distance that can be aligned by the shifting operation of the peak alignment process. As shown in Figure \ref{fig:case_studies_performance_1} (k), the trends of the other peaks we predicted are the same as the ground truth peaks.  
The last cubic test case MgAgF$_3$ is shown in Figure \ref{fig:case_studies_performance_1} (i). We find that due to fewer training samples contains elements Ag and Mg, the positions of the two highest peaks of MgAgF$_3$ (Figure \ref{fig:case_studies_performance_1} (l)) are predicted with larger offsets than previous examples. However, the positions of other peaks and intensities are still very close to the ground truth.

\begin{figure}[htpb!] 
    \begin{subfigure}[t]{0.28\textwidth}
        \includegraphics[width=\textwidth]{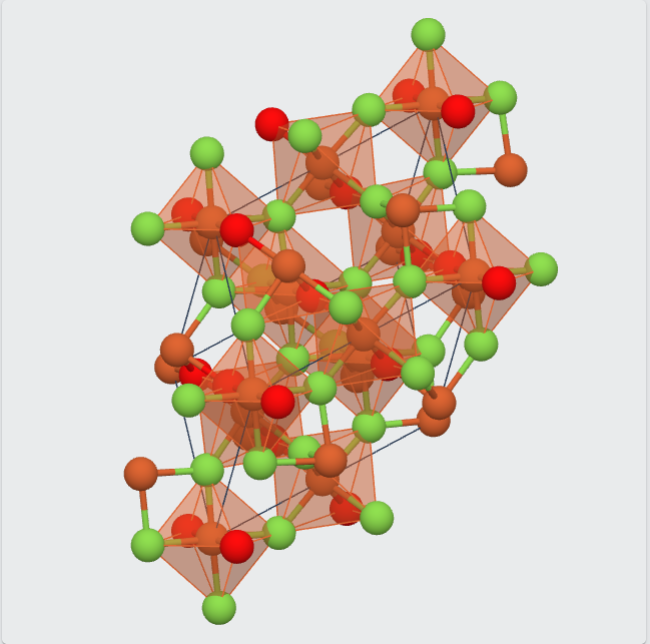}
        \caption{Fe$_3$(OF$_2$)$_2$}
        \vspace{-3pt}
    \end{subfigure}\hfill
    \begin{subfigure}[t]{0.28\textwidth}
        \includegraphics[width=\textwidth]{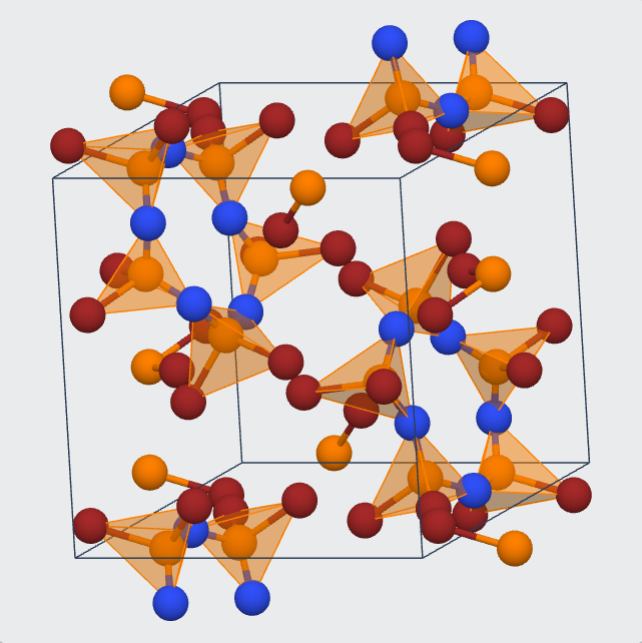}
        \caption{PBr$_2$N}
        \vspace{-3pt}
    \end{subfigure}\hfill
    \begin{subfigure}[t]{0.28\textwidth}
        \includegraphics[width=\textwidth]{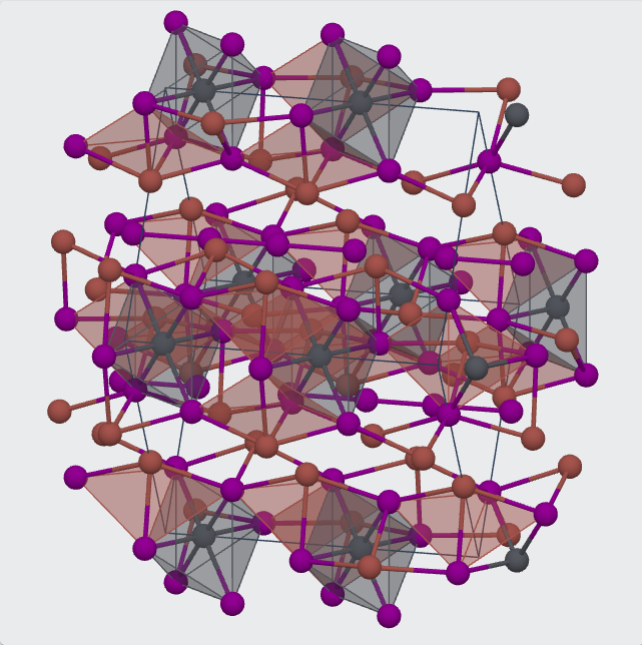} 
        \caption{Tl$_6$PbI$_10$}
        \vspace{-3pt}
    \end{subfigure}\hfill
    \vspace{1pt}
    \begin{subfigure}[t]{0.28\textwidth}
        \includegraphics[width=\textwidth]{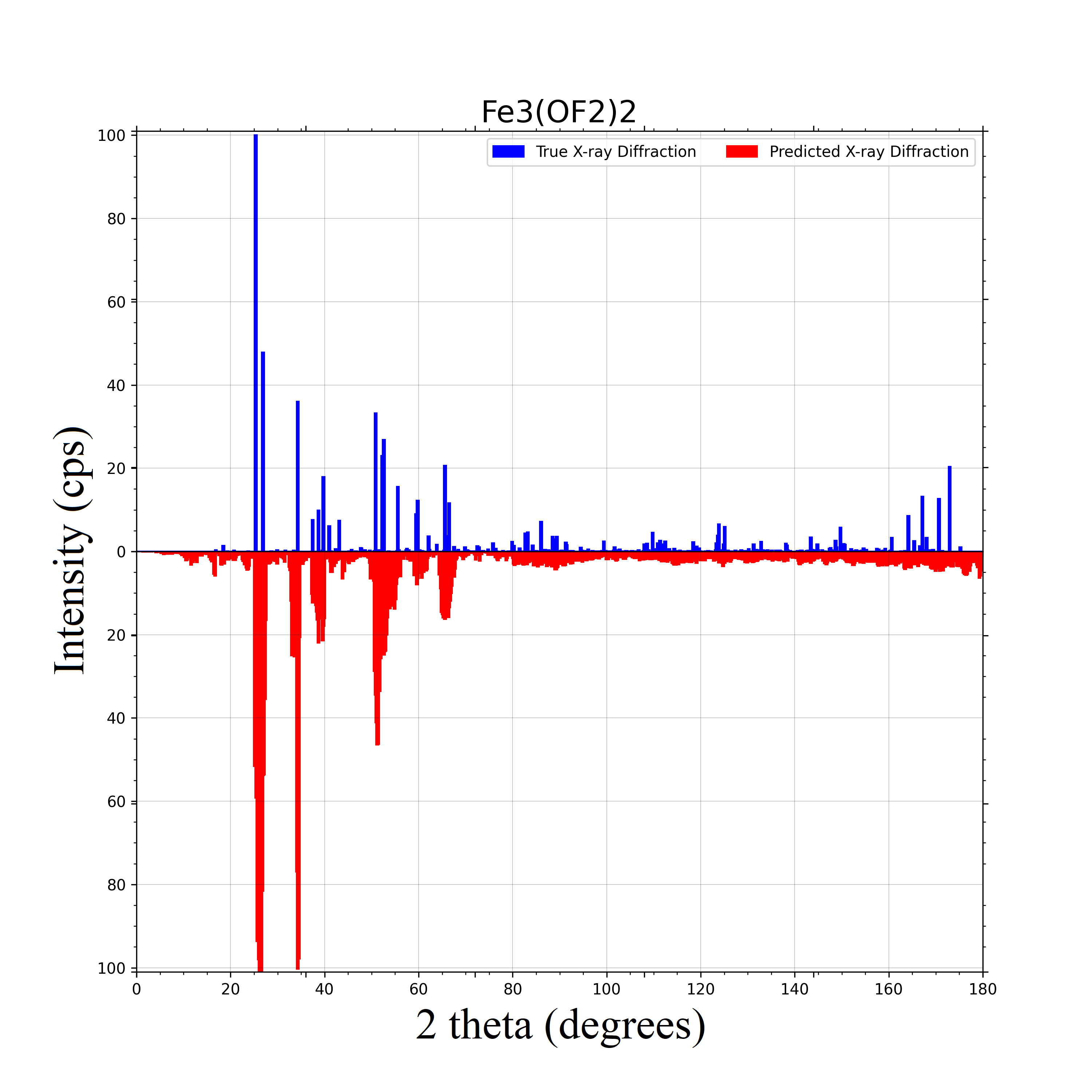}
        \caption{Fe$_3$(OF$_2$)$_2$ predicted XRD}
        \vspace{-3pt}
    \end{subfigure}\hfill    
    \begin{subfigure}[t]{0.28\textwidth}
        \includegraphics[width=\textwidth]{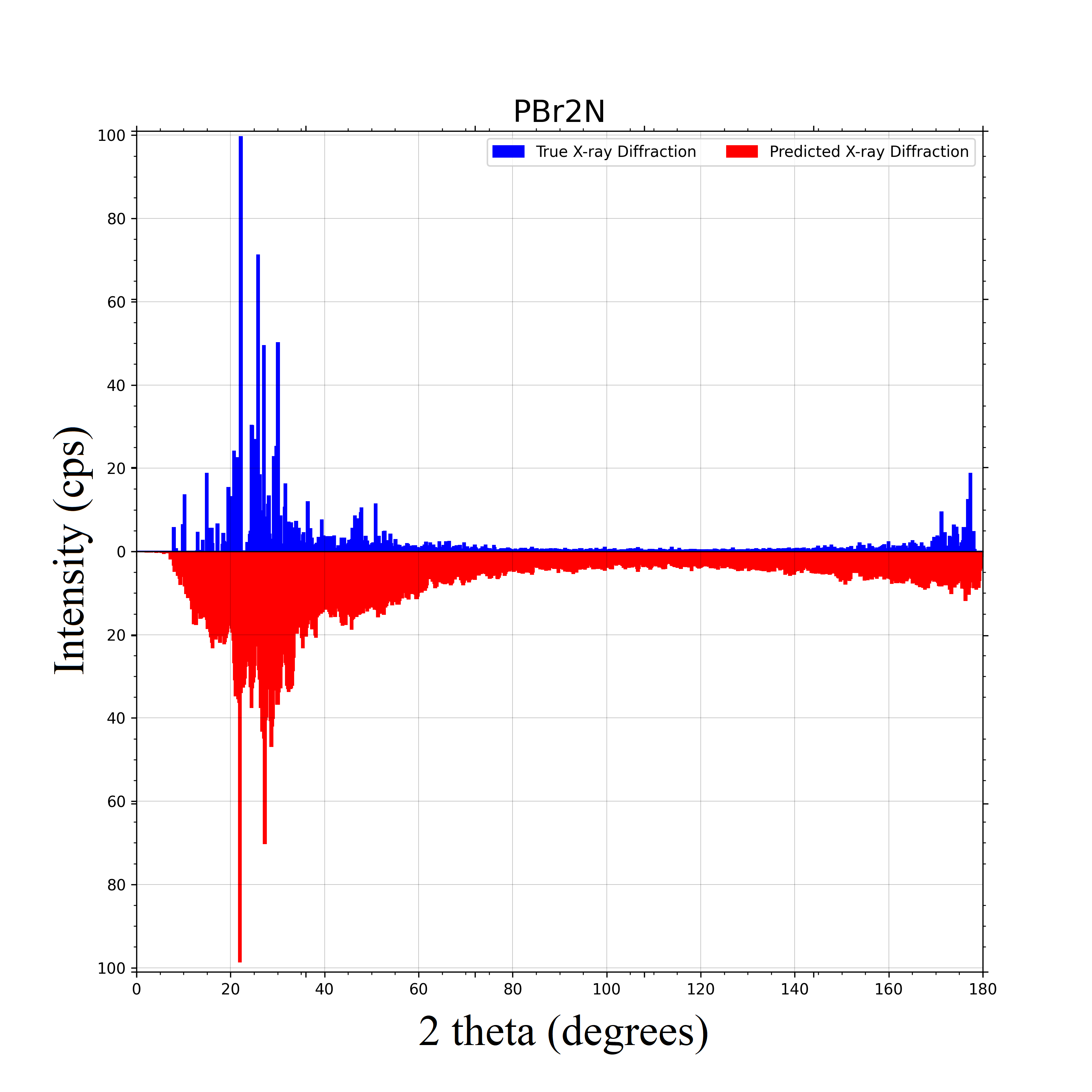}
        \caption{PBr$_2$N predicted XRD}
        \vspace{-3pt}
    \end{subfigure}\hfill
    \begin{subfigure}[t]{0.28\textwidth}
        \includegraphics[width=\textwidth]{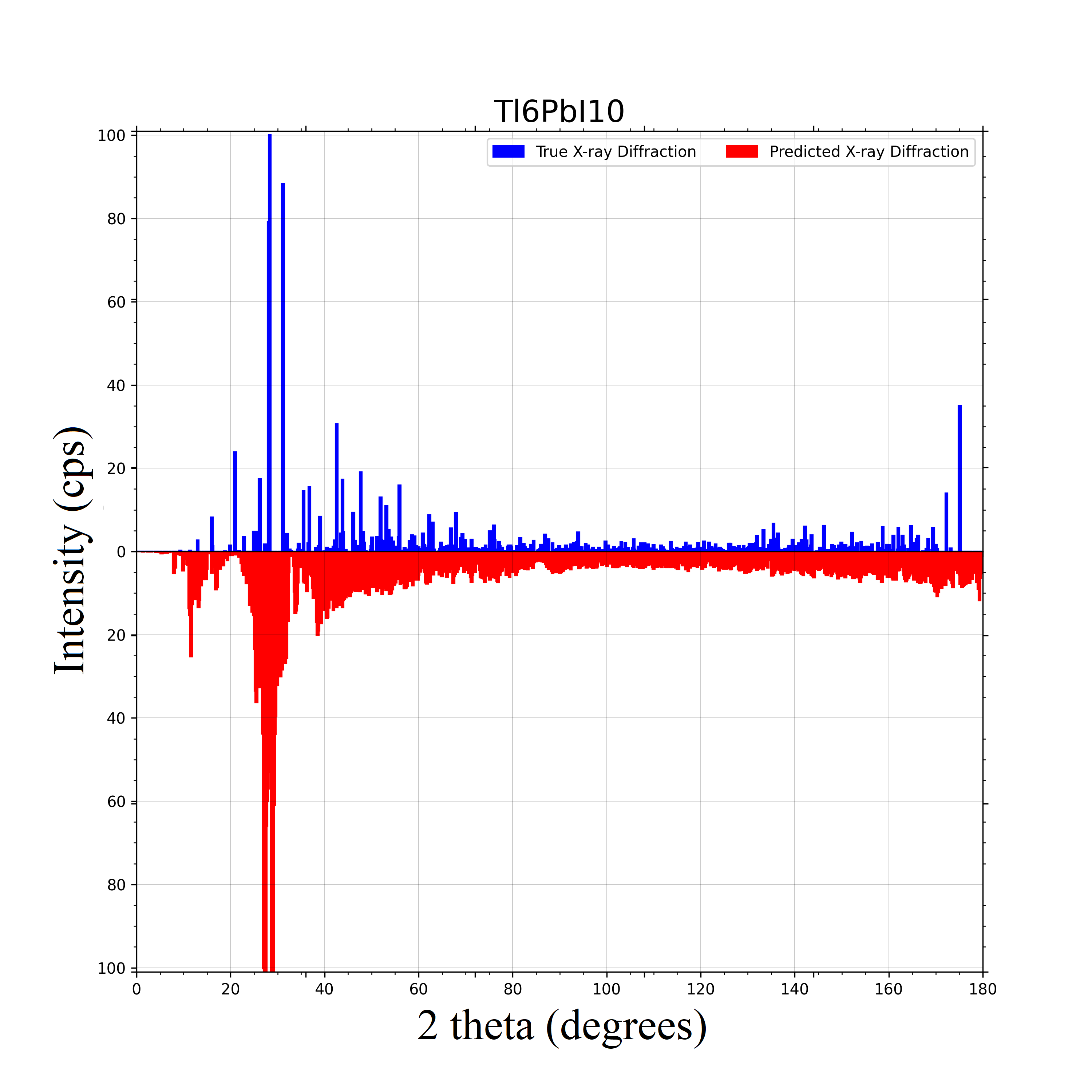}
        \caption{Tl$_6$PbI$_10$ predicted XRD}
        \vspace{-3pt}
    \end{subfigure}\hfill
    \begin{subfigure}[t]{0.28\textwidth}
        \includegraphics[width=\textwidth]{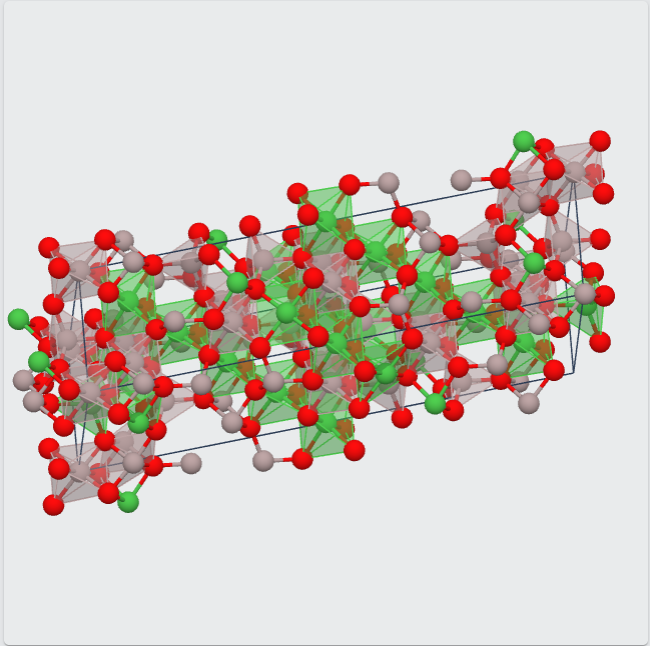}
        \caption{Al$_2$NiO$_4$}
        \vspace{-3pt}
    \end{subfigure}\hfill
    \begin{subfigure}[t]{0.28\textwidth}
        \includegraphics[width=\textwidth]{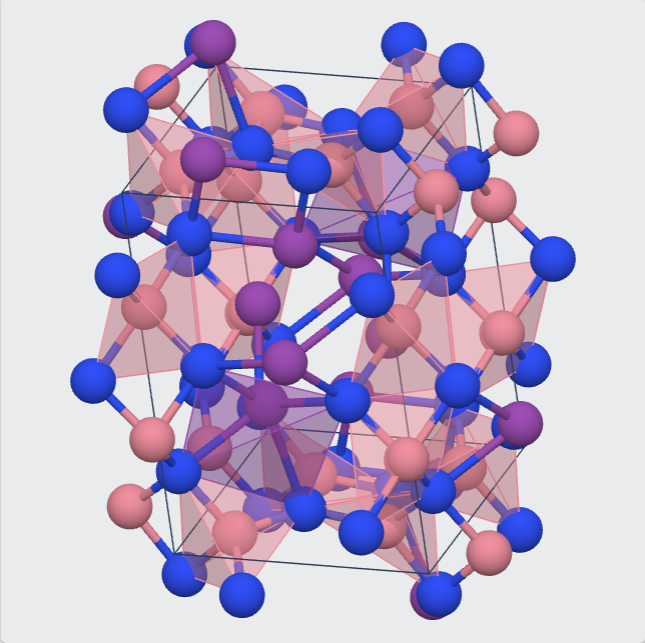}
        \caption{Co$_3$Bi$_3$N$_5$}
        \vspace{-3pt}
    \end{subfigure}\hfill
    \begin{subfigure}[t]{0.28\textwidth}
        \includegraphics[width=\textwidth]{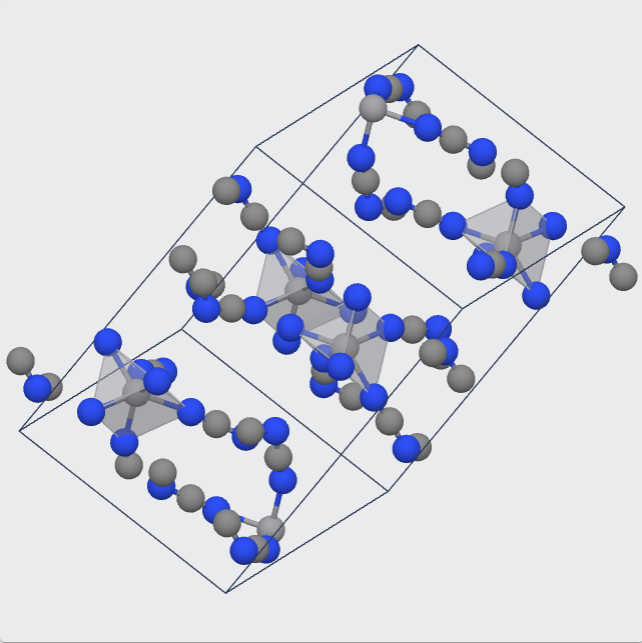}
        \caption{V(C$_2$N$_3$)$_3$}
        \vspace{-3pt}
    \end{subfigure}\hfill
    \begin{subfigure}[t]{0.28\textwidth}
        \includegraphics[width=\textwidth]{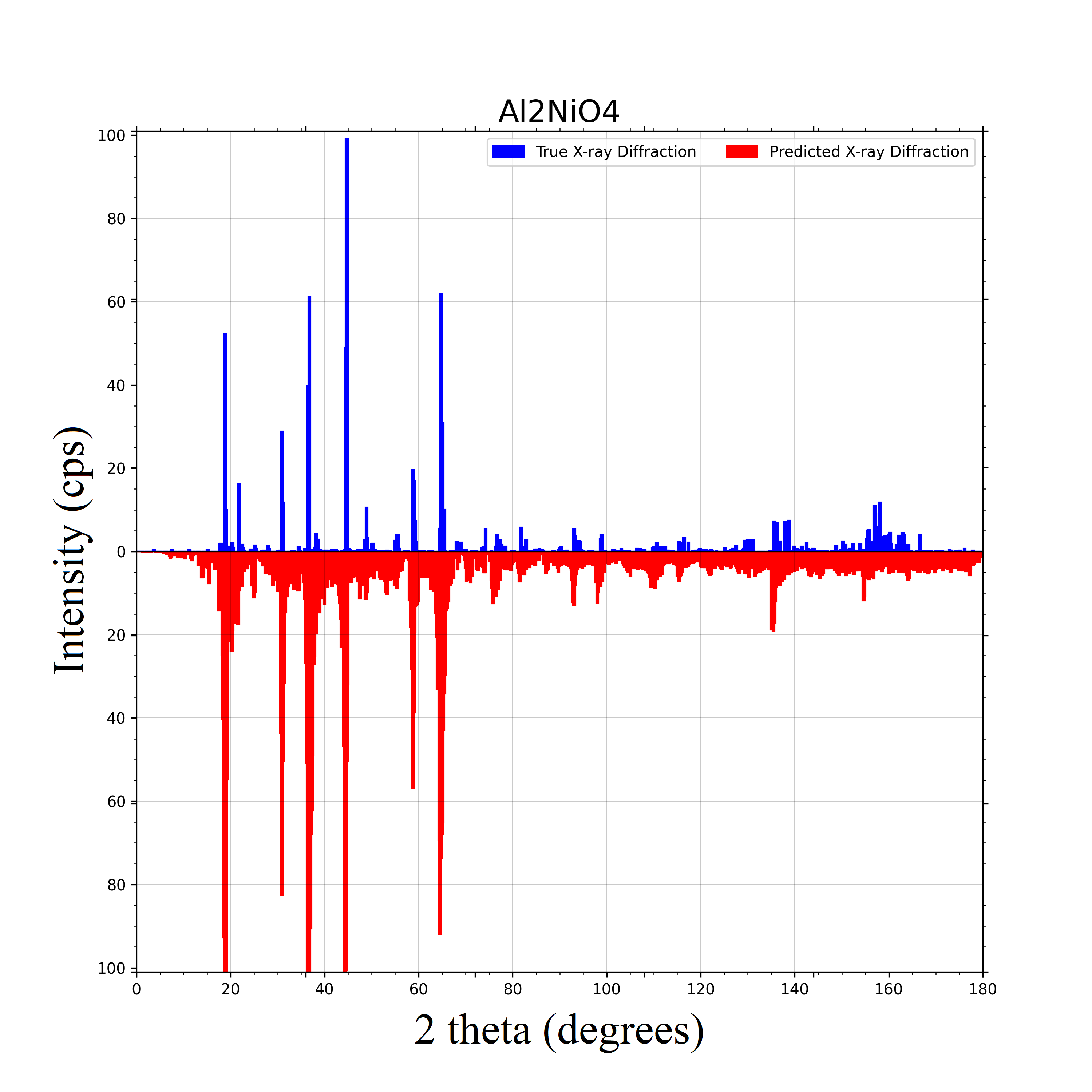}
        \caption{Al$_2$NiO$_4$ predicted XRD}
        \vspace{-3pt}
    \end{subfigure}\hfill    
    \begin{subfigure}[t]{0.28\textwidth}
        \includegraphics[width=\textwidth]{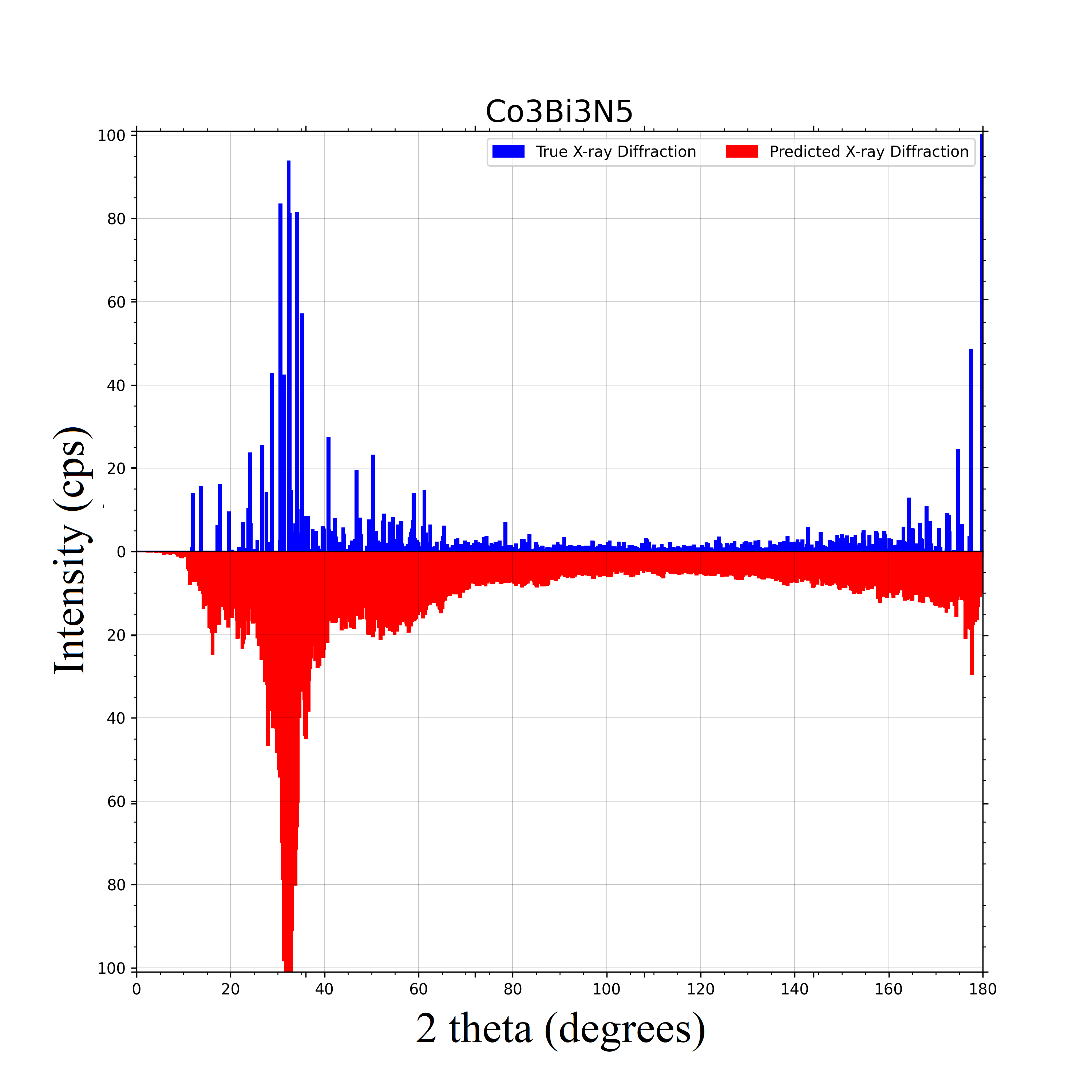}
        \caption{Co$_3$Bi$_3$N$_5$ predicted XRD}
        \vspace{-3pt}
    \end{subfigure}\hfill
    \begin{subfigure}[t]{0.28\textwidth}
        \includegraphics[width=\textwidth]{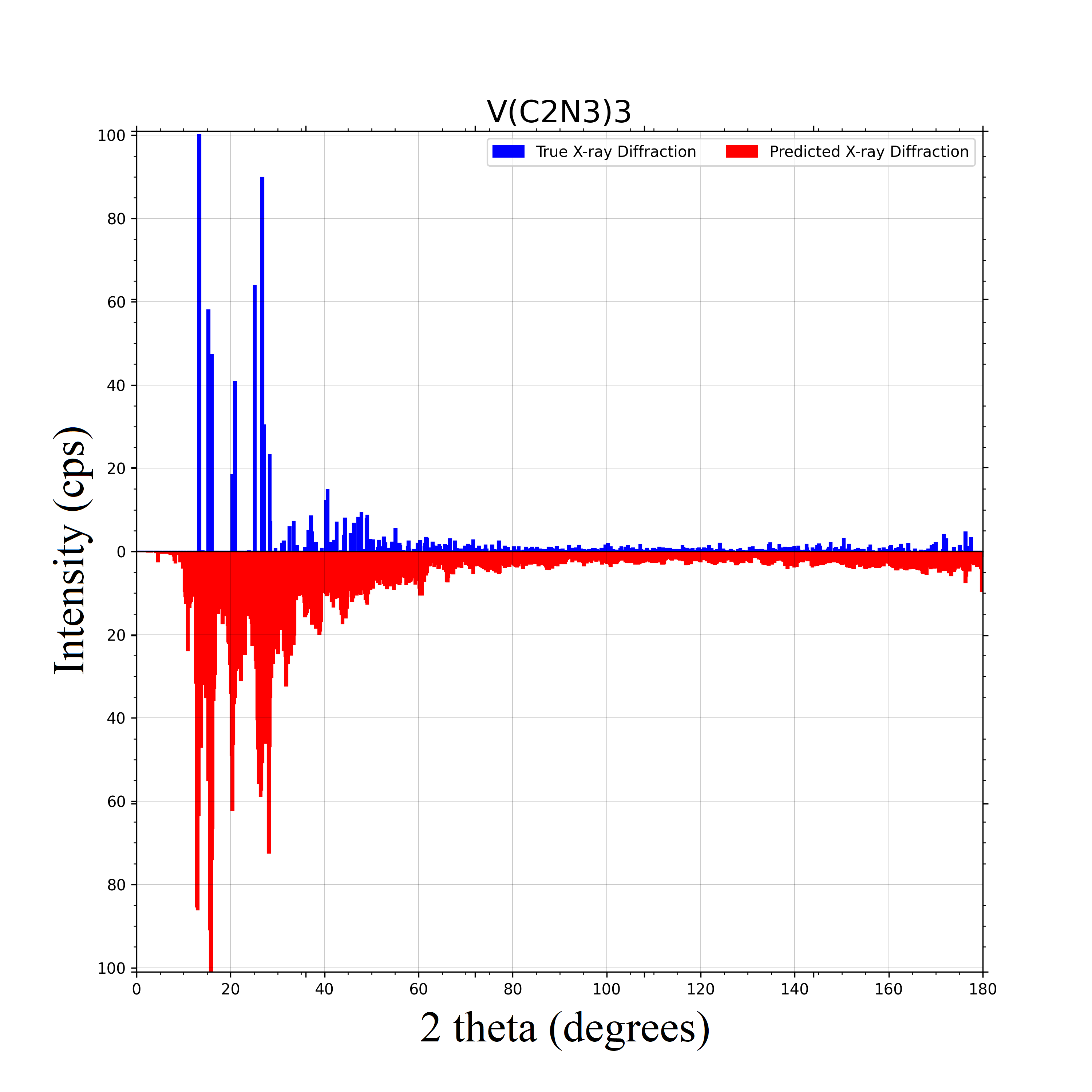}
        \caption{V(C$_2$N$_3$)$_3$ predicted XRD}
        \vspace{-3pt}
    \end{subfigure}\hfill   
    \caption{Prediction performance of DeepXRD. (a)(d): structure and predicted XRD of Fe$_3$(OF$_2$)$_2$; (b)(e): structure and predicted XRD of PBr$_2$N; (c)(f): structure and predicted XRD of Tl$_6$PbI$_10$ (g)(j): structure and predicted XRD of Al$_2$NiO$_4$; (h)(k): structure and predicted XRD of Co$_3$Bi$_3$N$_5$; (i)(l): structure and predicted XRD of V(C$_2$N$_3$)$_3$.}
    \label{fig:case_studies_performance_2}
\end{figure}

On the larger Ternary-XRD dataset, we also choose several test samples with different elements and crystal systems to evaluate our DeepXRD model's performance. Their predicted XRD (red color) and truth XRD spectra (blue color) are shown in Figure \ref{fig:case_studies_performance_2}. Comparing Figure \ref{fig:case_studies_performance_2} with Figure \ref{fig:case_studies_performance_1}, we can find that the peaks of these ternary materials are more complex than $ABC_3$. It is also found that our predicted XRD spectra are denser than those of $ABC_3$ dataset.

As shown in Figure \ref{fig:case_studies_performance_2} (a), Fe$_3$(OF$_2$)$_2$ is monoclinic. Its highest peak is located around 25 degrees, and most of the peaks are within the first 90 degrees. Our predicted XRD match almost all peaks of the first half and misses those peaks within the interval of 160 to 180 degrees again as shown in Figure \ref{fig:case_studies_performance_2} (d). The next test case is PBr$_2$N with triclinic structure as shown in Figure \ref{fig:case_studies_performance_2} (b). PBr$_2$N's higest peak position is around 20 degrees followed by a series of peaks with the intensities larger than 40 cps. Figure \ref{fig:case_studies_performance_2} (e) shows that almost all true peaks are located within the interval of 20 to 40 degrees, the same as the predicted ones. Figure \ref{fig:case_studies_performance_2} (c) shows the structure of Tl$_6$PbI$_10$, which has a trigonal crystal system. This crystal has three very close main peaks around 30 degrees (Figure \ref{fig:case_studies_performance_2} (f)). Our model accurately predicts the positions and intensities of all these peaks. Another monoclinic test case Al$_2$NiO$_4$ is shown in Figure \ref{fig:case_studies_performance_2} (g). Its peaks are discrete within the first 90 degrees as shown in Figure \ref{fig:case_studies_performance_2} (j). Our predicted XRD could match almost all the peak positions with intensity errors for only a few peaks. Figure \ref{fig:case_studies_performance_2} (h) shows a triclinic test case Co$_3$Bi$_3$N$_5$, which has a series of close high peaks and two single peaks around 180 degrees. Our predicted XRD matches the first series of peaks very well as shown in Figure \ref{fig:case_studies_performance_2} (k). The trends of the other predicted peaks are also similar to the true ones. 
The final case is the orthorhombic V(C$_2$N$_3$)$_3$ as shown in Figure \ref{fig:case_studies_performance_2} (i) with the peaks gathered between 0 and 30 degrees and our predicted XRD spectrum has the same distribution (Figure \ref{fig:case_studies_performance_2} (l)).




\subsection{Discussion}


From all three case studies discussed before, we show that for a given material formula, our DeepXRD model can predict its probable XRD spectrum only based on its composition. Even though the predicted peaks may not be at the exact positions compared to the ground truths, they are within the minor shifting range. For test cases with good performance, our model can find most peak positions and corresponding values but a few peaks may still not match, which may be caused by infrequent elemental combinations that only appear in limited times during model training. In machine learning studies, models trained with a larger dataset usually achieve better prediction performance. However, in our XRD study, the predicted XRD peaks of the test samples in the small $ABC_3$-XRD dataset match with their target XRD spectra better. This is due to the fact that the smaller ABC\textsubscript{3}-XRD dataset contains more similar compositions and structures compared to the samples of the Ternary-XRD dataset, which are much more diverse. During model training, we find that our model can easily overfit, which is probably due to the XRD spectrum data containing noise and being sensitive to its composition change, which means that even if there is one elemental change on the input formula, the corresponding XRD spectrum may change dramatically. If the training process focuses too much on the training set, the model tends to adapt to  the details or even noises within the XRDs of the training set, which makes the model to be not generalizable to the test samples. To deal with this overfitting problem, we add dropout layers in our model to control model overfitting and use early stopping to avoid our model's overfitting. We also find that the performance of our XRD prediction models may be significantly improved by designing a smoother loss function: instead of directly comparing the magnitudes at the sampling points, a loss function that allows a certain degree of peak shifting should lead to a smoother loss landscape so that similar compositions can have similar distribution despite some those peak shifts.


\section{Conclusion}

We propose a deep neural network-based model for predicting materials' XRD spectra given their composition only. These models can be used to conduct high-throughput screening of the almost infinite composition design space for structures with specific structural features or symmetry. When evaluated on two datasets with a more homogeneous $ABC_3$-XRD dataset and a larger Ternary-XRD dataset with more diverse structures, we show that our DeepXRD algorithm can make an accurate prediction of XRD spectra for a large category of material formulas. When we want to find materials with target XRD spectra, we can use the DeepXRD model as preliminary screening to narrow down the candidates. Based on the predicted XRD spectra, we may further estimate the material's structure. Based on our successful case studies, we believe that our DeepXRD model and its future variants are of great significance to be used for guiding the discovery of new materials.

\section{Contribution}
Conceptualization, J.H.; methodology, J.H. and R.D.; software, R.D., Y.Z., and J.H.; validation, R.D. and J.H.;  investigation, R.D., J.H., N.Fu, S. Omee, S. Dey, Q. Li, L. Wei; resources, J.H.; writing--original draft preparation, R.D. and J.H.; writing--review and editing, J.H; visualization, R.D.; supervision, J.H.;  funding acquisition, J.H.

\section{Acknowledgement}
Research reported in this work was supported in part by NSF under grant and 1940099 and 1905775 and by NSF SC EPSCoR Program under award number NSF Award OIA-1655740. The views, perspective, and content do not necessarily represent the official views of the SC EPSCoR Program nor those of the NSF.

\bibliography{references}
\bibliographystyle{unsrt}

\end{document}